\documentclass[twocolumn, times, tighten, appendixfloats]{aastex631}

\usepackage{natbib}
\usepackage{enumitem}
\usepackage{amsmath}
\usepackage{soul}
\usepackage{longtable}

\usepackage{graphicx}		
\usepackage{latexsym}		
\usepackage{bm}			
\usepackage[english]{babel}
\usepackage{color}		
\begin{document}

\title{JWST's PEARLS: Transients in the MACS J0416.1$-$2403 Field}

\author[0000-0001-7592-7714]{Haojing Yan} 
\affiliation{Department of Physics and Astronomy, University of Missouri, Columbia, MO\,65211, USA}

\author[0000-0003-3270-6844]{Zhiyuan Ma}
\affiliation{Department of Astronomy, University of Massachusetts, Amherst, MA 01003, USA}

\author[0000-0001-7957-6202]{Bangzheng Sun}
\affiliation{Department of Physics and Astronomy, University of Missouri, Columbia, MO\,65211, USA}

\author[0000-0001-7092-9374]{Lifan Wang}
\affiliation{George P. and Cynthia Woods Mitchell Institute for Fundamental Physics \& Astronomy, \\
Texas A. \& M. University, Department of Physics and Astronomy, 4242 TAMU, College Station, TX 77843, USA}

\author[0000-0003-3142-997X]{Patrick Kelly} 
\affiliation{School of Physics and Astronomy, University of Minnesota, 116
Church Street SE, Minneapolis, MN 55455, USA} 

\author[0000-0001-9065-3926]{Jos\'e M. Diego}
\affiliation{Instituto de Física de Cantabria (CSIC-UC), Avda. Los Castros s/n, 39005, Santander, Spain}

\author[0000-0003-3329-1337]{Seth H.~Cohen} 
\affiliation{School of Earth \& Space Exploration, Arizona State University, Tempe, AZ 85287-1404, USA}
\author[0000-0001-8156-6281]{Rogier A.~Windhorst} 

\affiliation{School of Earth \& Space Exploration, Arizona State University, Tempe, AZ 85287-1404, USA}
\affiliation{Department of Physics, Arizona State University, Tempe, AZ 85287-1504, USA}

\author[0000-0003-1268-5230]{Rolf A.~Jansen} 
\affiliation{School of Earth \& Space Exploration, Arizona State University, Tempe, AZ 85287-1404, USA}

\author[0000-0001-9440-8872]{Norman A.~Grogin} 
\affiliation{Space Telescope Science Institute, 3700 San Martin Drive, Baltimore, MD 21218, USA}

\author[0000-0002-0005-2631]{John F. Beacom}
\affiliation{Center for Cosmology and AstroParticle Physics (CCAPP), The Ohio State University, Columbus, OH 43210, USA}
\affiliation{Department of Physics, The Ohio State University, Columbus, OH 43210, USA}
\affiliation{Department of Astronomy, The Ohio State University, Columbus, OH 43210, USA}

\author[0000-0003-1949-7638]{Christopher J. Conselice} 
\affiliation{Jodrell Bank Centre for Astrophysics, Alan Turing Building, 
University of Manchester, Oxford Road, Manchester M13 9PL, UK}

\author[0000-0001-9491-7327]{Simon P. Driver} 
\affiliation{International Centre for Radio Astronomy Research (ICRAR) and the
International Space Centre (ISC), The University of Western Australia, M468,
35 Stirling Highway, Crawley, WA 6009, Australia}

\author[0000-0003-1625-8009]{Brenda Frye} 
\affiliation{Steward Observatory, University of Arizona, 933 N Cherry Ave, Tucson, AZ, 85721-0009, USA}

\author[0000-0001-7410-7669]{Dan Coe} 
\affiliation{Space Telescope Science Institute, 3700 San Martin Drive, Baltimore, MD 21218, USA}

\author[0000-0001-6434-7845]{Madeline A. Marshall} 
\affiliation{National Research Council of Canada, Herzberg Astronomy \&
Astrophysics Research Centre, 5071 West Saanich Road, Victoria, BC V9E 2E7, 
Canada}
\affiliation{ARC Centre of Excellence for All Sky Astrophysics in 3 Dimensions
(ASTRO 3D), Australia}

\author[0000-0002-6610-2048]{Anton Koekemoer}
\affiliation{Space Telescope Science Institute, 3700 San Martin Drive,
Baltimore, MD 21218, USA}

\author[0000-0001-9262-9997]{Christopher N. A. Willmer}  
\affiliation{Steward Observatory, University of Arizona, 933 N Cherry Ave, Tucson, AZ, 85721-0009, USA}

\author[0000-0003-0429-3579]{Aaron Robotham}
\affiliation{International Centre for Radio Astronomy Research (ICRAR) and the
International Space Centre (ISC), The University of Western Australia, M468,
35 Stirling Highway, Crawley, WA 6009, Australia}

\author[0000-0002-9816-1931]{Jordan C. J. D'Silva} 
\affiliation{International Centre for Radio Astronomy Research (ICRAR) and the
International Space Centre (ISC), The University of Western Australia, M468,
35 Stirling Highway, Crawley, WA 6009, Australia}

\author[0000-0002-7265-7920]{Jake Summers}
\affiliation{School of Earth \& Space Exploration, Arizona State University,
Tempe, AZ 85287-1404, USA}

\author[0000-0001-6342-9662]{Mario Nonino} 
\affiliation{INAF-Osservatorio Astronomico di Trieste, Via Bazzoni 2, 34124
Trieste, Italy} 

\author[0000-0003-3382-5941]{Nor Pirzkal} 
\affiliation{Space Telescope Science Institute, 3700 San Martin Drive, Baltimore, MD 21218, USA}

\author[0000-0003-0894-1588]{Russell E. Ryan, Jr.} 
\affiliation{Space Telescope Science Institute, 3700 San Martin Drive, Baltimore, MD 21218, USA}

\author[0000-0002-6150-833X]{Rafael {Ortiz~III}} 
\affiliation{School of Earth and Space Exploration, Arizona State University,
Tempe, AZ 85287-1404, USA}

\author[0000-0001-9052-9837]{Scott Tompkins} 
\affiliation{School of Earth and Space Exploration, Arizona State University,
Tempe, AZ 85287-1404, USA}

\author[0000-0003-0883-2226]{Rachana A. Bhatawdekar}
\affiliation{European Space Agency, ESA/ESTEC, Keplerlaan 1, 2201 AZ Noordwijk, NL}

\author[0000-0003-0202-0534]{Cheng Cheng}
\affiliation{Chinese Academy of Sciences South America Center for Astronomy, National Astronomical Observatories, CAS, Beijing 100101, China}

\author[0000-0002-0350-4488]{Adi Zitrin}
\affiliation{Physics Department, Ben-Gurion University of the Negev, P.O. Box 653, Beer-Sheva 8410501, Israel}

\author[0000-0002-9895-5758]{S. P. Willner}
\affiliation{Center for Astrophysics \textbar\ Harvard  \& Smithsonian, 60 Garden St., Cambridge, MA 02138, USA}

\begin{abstract}

   With its unprecedented sensitivity and spatial resolution, the James Webb 
Space Telescope (JWST) has opened a new window for time-domain discoveries in
the infrared. Here we report observations in the only field that has received
four epochs (spanning 126 days) of JWST NIRCam observations in Cycle 1. This 
field is towards MACS J0416.1$-$2403, which is a rich galaxy cluster at 
redshift $z=0.4$ and is one of the Hubble Frontier Fields. We have 
discovered 14 transients from these data. Twelve of these transients happened 
in three galaxies (with $z=0.94$, 1.01, and 2.091) crossing a lensing caustic 
of the cluster,and these transients are highly magnified by gravitational 
lensing. These 12 transients are likely of similar nature to those previously 
reported based on the Hubble Space Telescope (HST) data in this field, i.e., 
individual stars in the highly magnified arcs. However, these twelve could not 
have been found by HST because they are too red and too faint. The other two 
transients are associated with background galaxies ($z=2.205$ and 0.7093) that 
are only moderately magnified, and they are likely supernovae. They indicate a 
de-magnified supernova surface density, when monitored at a time cadence of a 
few months to a $\sim$3--4~$\mu$m survey limit of AB~$\sim 28.5$~mag,  
of $\sim$0.5~arcmin$^{-2}$ integrated to $z\approx 2$. This survey depth is 
beyond the capability of HST but can be easily reached by JWST. 

\end{abstract}

\section{Introduction}

    New capabilities in multi-messenger and time-domain astronomy will open 
outstanding vistas for discovery, as highlighted, for example, in the Decadal 
Survey on Astronomy and Astrophysics 2020 (Astro2020)
\footnote{\url{https://nap.nationalacademies.org/resource/26141/interactive/}}.
A core challenge is in localizing sources on the sky and in redshift, with 
the primary technique being searches for electromagnetic counterparts, which 
calls for observatories with the best flux sensitivity and angular
resolution possible.

    Until recently, one of the leading observatories for this purpose was the 
Hubble Space Telescope (HST), which had many successes. For example, it has 
found high-redshift supernovae, especially those of type Ia, which constrain 
cosmological models. In many cases, these observations were integrated with 
large, general-purpose extragalactic surveys 
\citep[e.g.,][]{Riess2004, Amanullah2010, Suzuki2012, Riess2018}. 
Observations of high-redshift clusters have been important for increasing 
yields \citep[e.g.,][]{Dawson2009, Hayden2021}, while observations of 
low-redshift clusters have led to 
the first discovery of a supernova that is gravitationally lensed into 
multiple images \citep{Kelly2015, Kelly2016}. The light curves of such 
supernovae, and in particular the time delay between images, provide a new 
route to measure the Hubble--Lema{\^i}tre constant $H_0$
\citep[][]{VF2018, Grillo2018, Grillo2020, Kelly2023a, Kelly2023b}.
For another example, a novel type of transient 
phenomena---caustic-crossing transients---has been identified through HST 
observations \citep[][]{Kelly2018, Rodney2018, Chen2019, Kaurov2019}. These are
individual stars in highly magnified background galaxies lying very close to 
the critical curve of the lensing cluster, which are further magnified---
temporarily---by intracluster stars that act as microlenses. These transients 
have provided a completely unexpected method to study individual stars at
cosmological distances.

    The advent of the James Webb Space Telescope (JWST) has brought 
dramatically better opportunities because of its more than an order of 
magnitude sensitivity
increase relative to HST\null. The Prime Extragalactic Areas for 
Reionization and Lensing Science program \citep[PEARLS;][]{Windhorst2023}, 
one of the programs under the JWST Interdisciplinary Scientists' Guaranteed 
Time Observations (GTO), has a major time-domain science component. One of its
fields is MACS J0416.1$-$2403 (hereafter M0416), which is a lensing cluster 
at $z\sim 0.4$ and one of the Hubble Frontier Fields \citep[HFF:][]{Lotz2017}. 
The redshift of M0416 as a whole is still uncertain at the level of 
$\pm 0.003$, mainly because this is a merging cluster whose 
sub-components might have large peculiar motions. Spectroscopic campaigns on 
this cluster are described by \citet[][]{Balestra2016, Caminha2017, 
Vanzella2021, Bergamini2021}. We adopt $z=0.397$ as the fiducial redshift of 
this cluster.
HST has previously revealed several caustic-crossing transients near two 
caustic-straddling arcs in M0416. \citet[][]{Rodney2018} discovered two fast 
transients in an arc 
identified at $z=1.0054$ \citep[][]{Caminha2017,Rodney2018},
which was nicknamed 
``Spock'' by the authors. The transient sources themselves are consistent with 
being supergiant stars with temperatures between $3500$~K and $18000$~K 
residing in the strongly lensed galaxy that constitutes the Spock arc
\citep{Diego2023a}. \citet[][]{Chen2019} and \citet[][]{Kaurov2019} 
found a transient of similar nature in another arc 
identified at $z=0.94$ \citep[][]{Hoag2016, Caminha2017},
which is named ``Warhol.'' In addition, the ultra-deep, UV-to-visible HST 
program  ``Flashlights'' detected two high-significance caustic transients in 
the Spock arc and four in the Warhol arc \citep{Kelly2022}. Highly lensed 
regions such as these are expected to produce caustic transients continually.

    To take advantage of opportunities enabled by JWST, PEARLS
incorporated three epochs of NIRCam observations of M0416. We 
expected these data to be particularly powerful for detecting red supergiant 
stars at  $z\gtrsim 1$ because red supergiants are bright at 
$\lambda \gtrsim 2$~$\mu$m. The design was also motivated by the possibility 
of detecting individual Population III stars through caustic transits at 
$z>7$ \citep[][]{Windhorst2018}. Another JWST GTO program, the CAnadian 
NIRISS Unbiased Cluster Survey \citep[CANUCS;][]{Willott2022}, also observed M0416 with NIRCam in a 
separate epoch. All these data have 
been taken, making M0416 the only field in JWST Cycle~1 that has four 
epochs of NIRCam observations. This makes M0416  the best region in 
the sky to date for studying infrared transients.

    This paper reports a transient search using the unique 4-epoch data. The search
has gone beyond the aforementioned two arcs, as we also intend to assess the  
general infrared transient rate in less magnified regions at depths that have 
never been probed before. This paper is the first in a series on this subject 
and presents an overview of the transients found in this field. The paper is 
organized as follows. The NIRCam observations and data are described in 
Section~2. The transient search is detailed in Section~3. Section~4 discusses the 
transient, and Section~5 summarizes results.  All 
magnitudes are in the AB system, and all coordinates are in the ICRS frame 
(equinox 2000). 

\section{Observations and Data}

   The four epochs of NIRCam observations all used the same eight bands, namely,
F090W, F115W, F150W, and F200W in the ``short wavelength'' (SW) channel and
F277W, F356W, F410M, and F444W in the ``long wavelength'' (LW) channel. The native NIRCam pixel scales are 0\farcs031~pix$^{-1}$ in the
SW channel and 0\farcs063~pix$^{-1}$ in the LW channel. As the SW channel is 
made up of four detectors, the observations used the \texttt{INTRAMODULEBOX} 
dithers to cover the gaps. The PEARLS observations adopted the 
\texttt{MEDIUM8} readout pattern with ``up-the-ramp'' fitting to determine the 
count rate, while those of CANUCS used a combination of the \texttt{SHALLOW4} 
and \texttt{DEEP8} patterns. The total exposure times, dates of observation, 
and 5~$\sigma$ depths of these observations are summarized in Table~\ref{tbl:obs}. 

\begin{table}[hbt!]
\caption{M0416 NIRCam Observation Summary}
\label{tbl:obs}
\centering
{\footnotesize
\begin{tabular}{cccc} \hline\hline
 Epoch  & Filter & Exptime &  Depth \\
 Start UT && (s)& 5$\sigma$\\ \hline
 Ep1  & F090W & 3779.343 & 28.45 \\ 
 2022 Oct 7 08:06:05  & F115W & 3779.343 & 28.48 \\ 
 (0.0 days) & F150W & 2920.401 & 28.48 \\ 
 ($\rm PA=293\arcdeg$) & F200W & 2920.401 & 28.69 \\ 
 & F277W & 2920.401 & 29.83 \\ 
 & F356W & 2920.401 & 29.91 \\
 & F410M & 3779.343 & 29.38 \\ 
 & F444W & 3779.343 & 29.58 \\ 
 \hline
 Ep2  & F090W & 3779.343 & 28.49 \\ 
 2022 Dec 29 16:00:36  & F115W & 3779.343 & 28.51 \\ 
 (83.4 days) & F150W & 2920.401 & 28.51 \\ 
 ($\rm PA=33\arcdeg$)& F200W & 2920.401 & 28.72 \\ 
 & F277W & 2920.401 & 29.88 \\ 
 & F356W & 2920.401 & 29.93 \\
 & F410M & 3779.343 & 29.39 \\ 
 & F444W & 3779.343 & 29.59 \\ 
 \hline
 Ec  & F090W & 6399.115 & 28.63 \\ 
 2023 Jan 11 20:24:42 & F115W & 6399.115 & 28.64 \\ 
 (96.7 days) & F150W & 6399.115 & 28.82 \\ 
 ($\rm PA=49\arcdeg$)& F200W & 6399.115 & 29.02 \\ 
 & F277W & 6399.115 & 30.14 \\ 
 & F356W & 6399.115 & 30.19 \\
 & F410M & 6399.115 & 29.50 \\ 
 & F444W & 6399.115 & 29.70 \\ 
 \hline
 Ep3 &  F090W & 3779.343 & 28.48 \\ 
 2023 Feb 10 09:12:32 &  F115W & 3349.872 & 28.41 \\ 
 (126.1 days) & F150W & 2920.401 & 28.49 \\ 
 ($\rm PA=71\arcdeg$)& F200W & 2920.401 & 28.69 \\ 
 & F277W & 2920.401 & 29.84 \\ 
 & F356W & 2920.401 & 29.86 \\
 & F410M & 3349.872 & 29.21 \\ 
 & F444W & 3779.343 & 29.47 \\ 
 \hline
\end{tabular}
}
\raggedright
\tablecomments{The 5$\sigma$ depths (in AB magnitudes ) are measured from the
RMS map within circular apertures of 0\farcs2 radius. PA is the angle of the 
detector $y$ axis projected on the sky in degrees east from north.}
\end{table}

    NIRCam has two nearly identical modules (``A'' and ``B'') that subtend two
adjacent, square fields. As the spatial orientations of the JWST instruments 
vary in time on an annual basis, these two fields cannot both be on the same 
region in the sky within a year. For this reason, all four epochs of NIRCam 
observations were designed to center the B module on the cluster, leaving the
A module mapping different regions in the flanking area. This transient
study uses only the module B data because only these are spatially
overlapped.

\begin{figure*}[t]
  \epsscale{1.0}
  \plotone{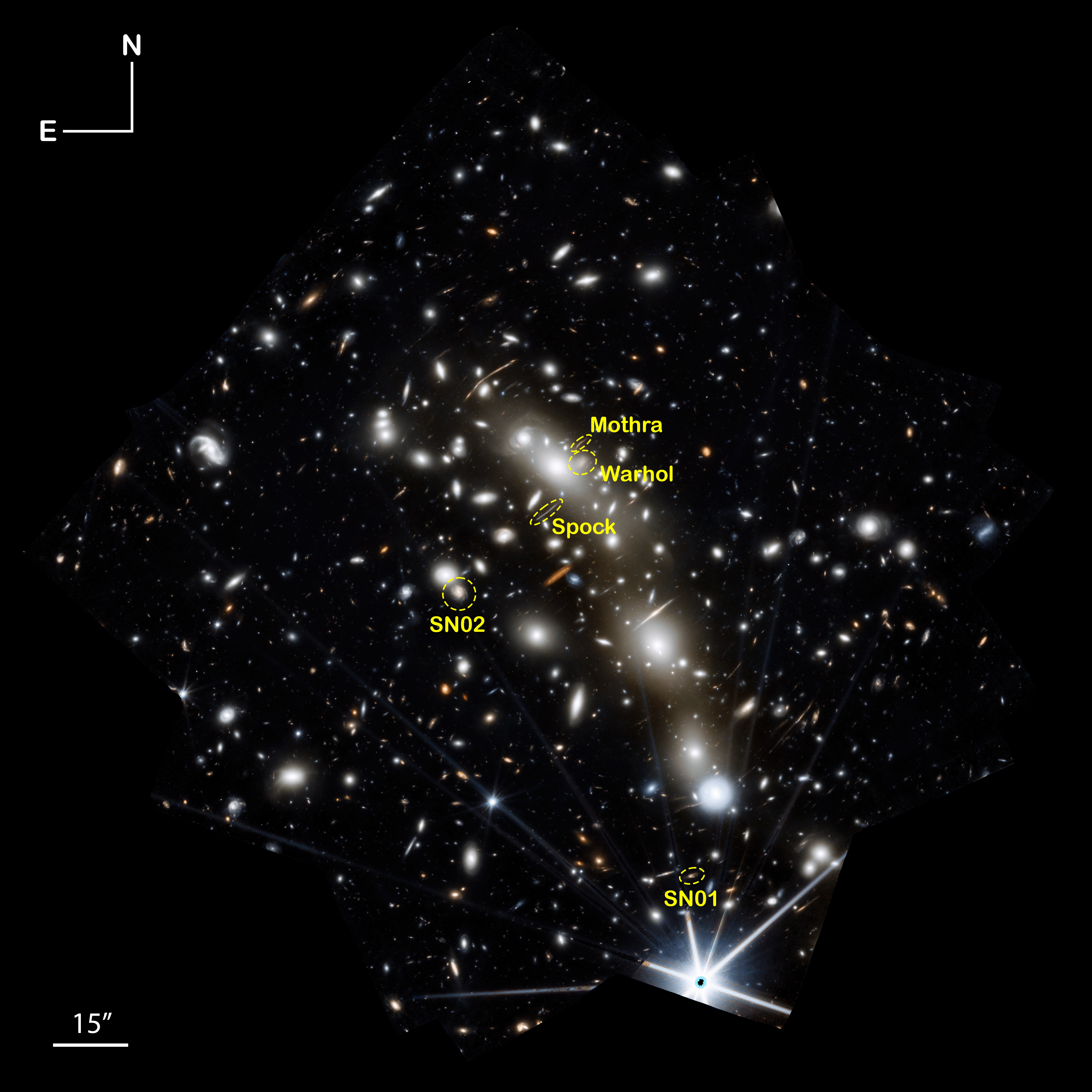}
    \caption{Color composite image of M0416 using the data from four epochs of 
NIRCam observations as described in the text. The color scheme is
F090W+F115W+F150W as blue, F200+F277W as green, and F356W+F410M+F444W as red.
The regions where the transients are found are also marked.
    }  
    \label{fig:overview}
\end{figure*}

   The data were retrieved from the Mikulski Archive for Space Telescopes
(MAST). Reduction started from the so-called Stage~1 ``uncal'' products, which 
are the single exposures from the standard JWST data reduction pipeline 
\citep[][]{bushouse_howard_2023}
after Level~1b processing.
We further processed these products using the version 
1.9.4 pipeline in the context of \texttt{jwst\_1063.pmap}. A few changes and 
augmentations were made to the pipeline to improve the reduction quality; most 
importantly, these included enabling the use of an external reference catalog 
for image alignment and implementing a better background estimate for the final 
stacking. The astrometry of each single exposure was calibrated using
the public HFF products
\footnote{\url{https://archive.stsci.edu/prepds/frontier/macs0416.html}}.
These single images were projected onto the same grid and
were stacked in each band and in each epoch. We produced two versions of 
stacks, one at the pixel scale of 0\farcs06 (hereafter the ``60mas'' version)
and the other at the scale of 0\farcs03 (the ``30mas'' version), to best match
the native pixel scales in the LW and the SW channels, respectively. The 
mosaics are in surface brightness units of MJy~sr$^{-1}$. The AB magnitude 
zeropoints are 26.581 and 28.087 for the 60mas and the 30mas stacks,
respectively. Figure~\ref{fig:overview} shows a composite color image using 
the data from all four epochs. For convenience, we refer to the
three epochs from the PEARLS program as Ep1, Ep2, and Ep3, respectively, where
``p'' stands for ``PEARLS\null.'' The epoch from the CANUCS program, which was 
between Ep2 and Ep3, is referred to as Ec (``c'' for ``CANUCS'').

\section{Transient Discoveries}

\subsection{Search method}

   The search for transients was done in the usual way by detecting positive 
peaks on difference images between epochs. Thanks to the excellent image 
alignment and stable image quality, we were able to form the difference images 
by direct subtraction, for which we used the 60mas images. Due to the intense 
human labor in the visual inspection step (see below), this work is limited to 
these pairs of difference images: Ep1~$-$~Ep2 for the search of decaying 
sources in Ep1, Ep2~$-$~Ep1 for new sources appearing in Ep2, and Ep3~$-$~Ep1 
for new sources appearing in Ep3. Ec was not used to initiate transient 
searches, as it was only 13.3 days after Ep2; however, it was used when 
studying the light curves of the identified transients. When building a 
difference image, its ``root mean square'' (RMS) map was also constructed by 
adding the RMS maps of the parent images in quadrature. As compared to the SW
images, the LW images are less affected by defects and therefore their 
difference images are cosmetically cleaner. We  used the F356W band as 
the basis of our search because its data are the deepest.

   The initial transient search was to run SExtractor 
\citep{Bertin1996} on the 60mas F356W difference images. The 
RMS maps were used in the process to estimate the signal-to-noise ratios (S/N) 
of the peaks in the difference images. Only the peaks that have 
$\rm{S/N}\geq 5$ were further considered. This left thousands of peaks in each
difference image, which were then visually inspected. Not surprisingly, the vast
majority of these peaks are not transient objects but are residuals caused by
imperfect subtraction of bright stars and galaxies for two reasons: (1) the
values of the pixels occupied by bright objects fluctuate over different epochs
because of Poisson noise, which often results in positive peaks accompanied by 
negative peaks in a difference image; (2) the position angle of the point 
spread function (PSF) is different in different epochs owing to the different 
field orientations. This leads to spurious sources around bright objects that 
appear in different positions in different epochs. 

After the initial visual inspection, only a few tens of transient candidates survived.
    To ensure their reliability, we further required that the selected 
transients should be detected at $\rm{S/N}\geq 5$ in the difference image of at 
least one more band in addition to F356W\null. Applying this requirement gave a 
total of 14 robust transients.

\subsection{Descriptions of the transients and their photometry}

   Our sample includes seven transients in the Warhol region, four in the Spock
region, one in yet another arc, and two in other regions. Their locations are 
indicated in Figure \ref{fig:overview}. These objects and their photometry are 
described below. Their magnification factors based on the lens model of 
\citet[][adopting their 68\%\ confidence level intervals]{Bergamini2023} are
also quoted along with the photometry.

In most cases, these sources are embedded in a highly 
non-uniform background and/or are affected by contamination from nearby 
objects, and PSF fitting had to be used to obtain reliable photometry. For this
purpose, the 30mas images are more appropriate. To be consistent, we used PSF 
fitting for all objects on the 30mas images, and the detailed process is 
explained in the Appendix.

\begin{table*}[hbt!]
\centering
\caption{Catalog of transients in the Warhol region.}
\label{tbl:WR_phot}
\resizebox{\textwidth}{!}{
\begin{tabular}{ccccccccccccc}\hline
 & R.A. & Decl. & Epoch & F090W & F115W & F150W & F200W & F277W & F356W & F410M & F444W & $\mu$ \\ \hline
 {\bf D21-W1} & 64.03695 & $-$24.06725 & Ep1 & 29.74\textpm0.44 & 28.89\textpm0.13 & 28.15\textpm0.06 & 27.90\textpm0.05 & 27.65\textpm0.07 & 27.97\textpm0.11 & 28.05\textpm0.12 & 28.46\textpm0.13 & 26.6\\ 
 & & & Ep2 & 29.41\textpm0.19 & 28.68\textpm0.10 & 27.92\textpm0.05 & 27.34\textpm0.04 & 27.10\textpm0.05 & 27.19\textpm0.05 & 27.42\textpm0.06 & 27.56\textpm0.06 & \\ 
 & & & Ec & 29.09\textpm0.22 & 28.96\textpm0.12 & 28.28\textpm0.06 & 27.78\textpm0.04 & 27.61\textpm0.06 & 27.65\textpm0.06 & 27.79\textpm0.07 & 28.08\textpm0.08 & \\ 
 & & & Ep3 & 29.53\textpm0.34 & 29.36\textpm0.35 & 29.03\textpm0.14 & 29.00\textpm0.16 & 29.10\textpm0.30 & $>$29.93* & $>$29.51* & $>$29.73* & \\ \hline
  
 {\bf D21-W2} & 64.03674 & $-$24.06725 & Ep2 & 29.43\textpm0.21 & $>$29.28 & 28.88\textpm0.16 & 28.99\textpm0.17 & 28.37\textpm0.10 & 28.11\textpm0.09 & 28.08\textpm0.11 & 28.83\textpm0.15 & 904.2 \\ 
 & & & Ec & 29.26\textpm0.17 & 29.07\textpm0.15 & 29.21\textpm0.20 & 29.49\textpm0.25 & 28.53\textpm0.11 & 28.40\textpm0.12 & 28.25\textpm0.15 & 28.78\textpm0.18 & \\
 & & & Ep3 & 28.33\textpm0.08 & 28.31\textpm0.08 & 28.55\textpm0.11 & 28.89\textpm0.16 & 29.19\textpm0.23 & 28.67\textpm0.16 & 28.72\textpm0.20 & 29.09\textpm0.21 & \\ \hline
 
 {\bf D21-W3} & 64.03665 & $-$24.06728 & Ep2 & $>$29.31 & $>$29.28 & 29.72\textpm0.32 & 28.88\textpm0.14 & 28.67\textpm0.16 & 28.42\textpm0.12 & 28.69\textpm0.26 & 29.10\textpm0.23 & 331.3 \\
 & & & Ec & $>$29.40 & $>$29.38 & 30.13\textpm0.46 & 29.78\textpm0.33 & 29.09\textpm0.23 & 28.66\textpm0.14 & 28.59\textpm0.21 & 28.77\textpm0.17 & \\ \hline
 
  {\bf Dc2-W4} & 64.03655 & $-$24.06731 & Ec & $>$28.89* & $>$28.87* & 29.00\textpm0.19 & 28.18\textpm0.11 & 27.38\textpm0.07 & 27.04\textpm0.06 & 27.28\textpm0.06 & 27.36\textpm0.07 & 1188.6 \\ \hline
  
  {\bf D31-W5} & 64.03668 & $-$24.06732 & Ep3 & $>$29.32 & $>$29.29 & $>$29.24 & 29.28\textpm0.22 & 28.75\textpm0.18 & 28.62\textpm0.16 & 28.99\textpm0.31 & 29.62\textpm0.39 & 83.0 \\ \hline 
  
  {\bf D31-W6} & 64.03654 & $-$24.06744 & Ep3 & $>$29.32 & $>$29.29 & 29.13\textpm0.17 & 28.56\textpm0.09 & 28.32\textpm0.13 & 28.16\textpm0.10 & 28.17\textpm0.16 & 28.07\textpm0.11 & 62.8 \\ \hline
  
  {\bf D31-W7} & 64.03650 & $-$24.06736 & Ep3 & $>$29.32 & 29.86\textpm0.34 & 29.71\textpm0.30 & 28.78\textpm0.16 & 28.50\textpm0.15 & 28.31\textpm0.12 & 28.64\textpm0.20 & 28.77\textpm0.18 & 365.9 \\ \hline
\end{tabular}
}
\raggedright
\tablecomments{The R.A.\ and Decl.\ coordinates are in decimal degrees for the 
J2000.0 equinox. The magnitudes are based on the PSF-fitting results. The upper 
limits are measured within 11$\times$11 pixels (to match the size of the PSF-fitting area) centered at the source location on the RMS maps. The limits 
labeled with $\ast$ are 5$\sigma$ upper limits (appropriate for the 
measurements in the original image where the background is high and 
non-uniform), otherwise 2$\sigma$ (more appropriate for the measurements in the 
difference images where the background is largely subtracted). 
The quoted magnification factors $\mu$ are the 68\%\ confidence level intervals
from \citet[][]{Bergamini2023}.}
\end{table*}

\subsubsection{Transients in the Warhol region}

\begin{figure*}[t]
  \epsscale{1.2}
  \plotone{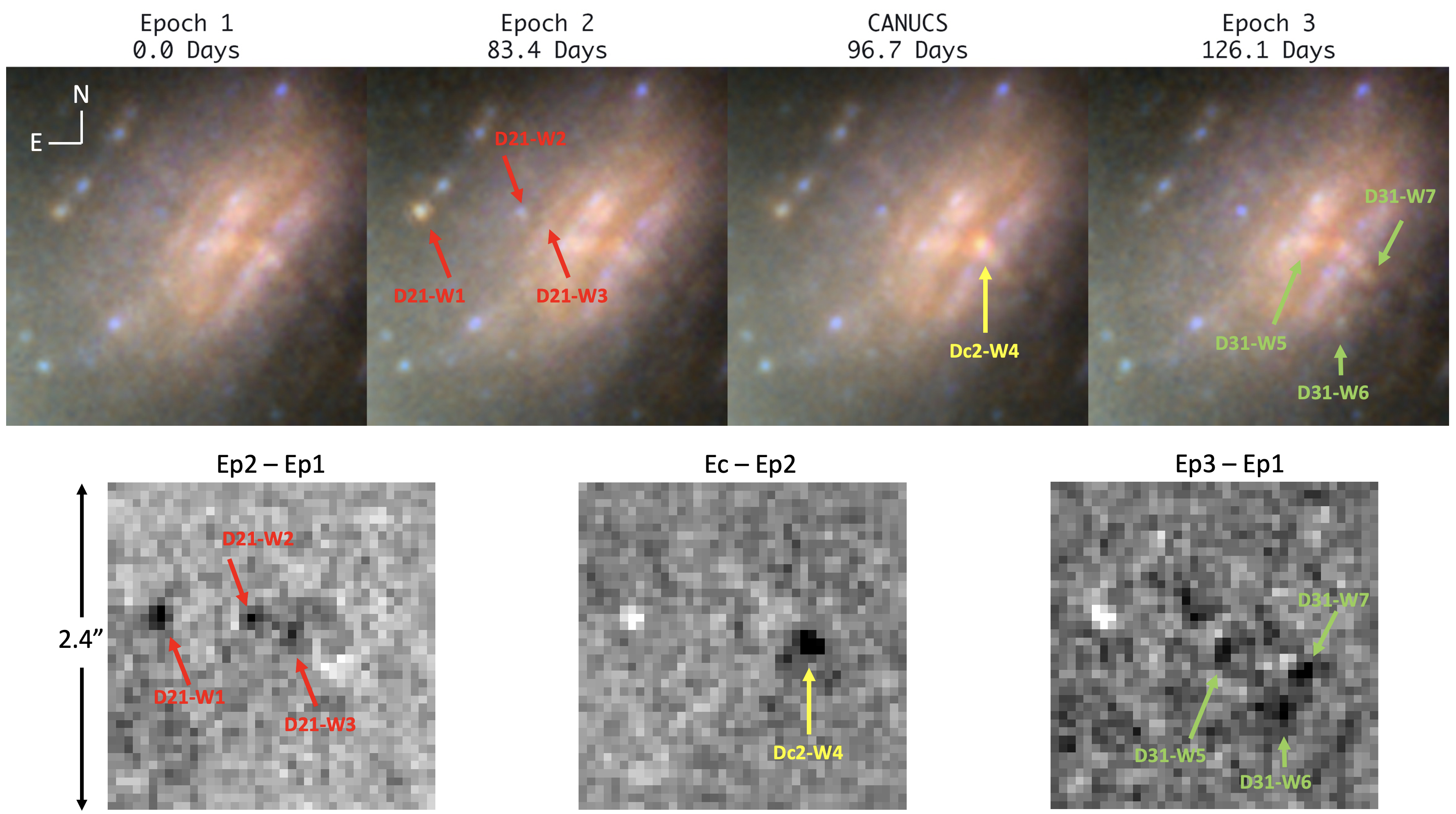}
    \caption{Locations of the transients found in the Warhol region. The upper
panel shows the color images  of this region in the four epochs, while the 
lower panels show the inverted F356W difference images of the same region  
between relevant epochs that led to the discovery of these transients as 
labeled. All these images are 2\farcs4 on a side and are oriented north-up 
and east-left.}
    \label{fig:WR}
\end{figure*}

\begin{figure*}[t]
  \epsscale{0.9}
  \plotone{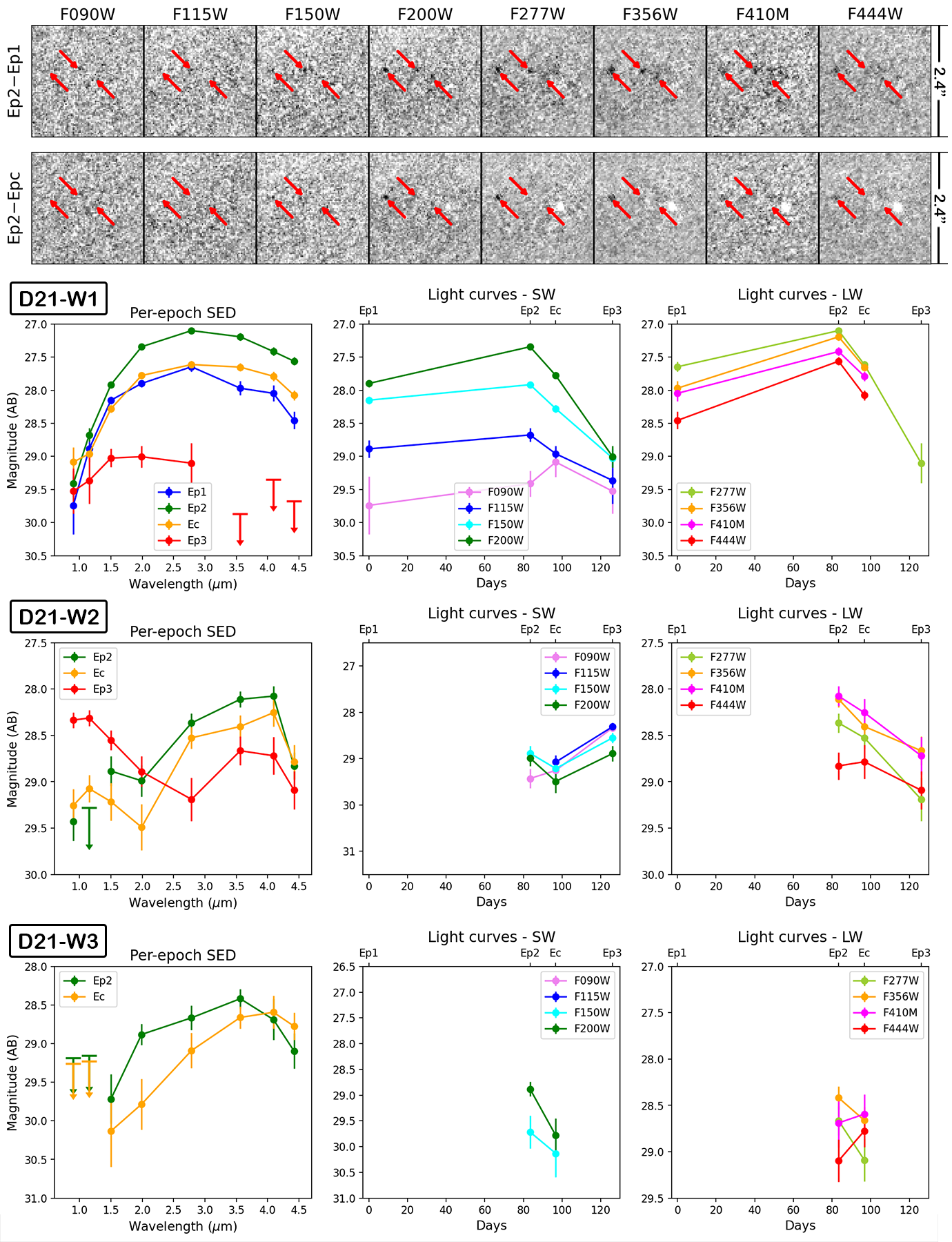}
    \caption{(top two rows) Transients D21-W1, D21-W2, and D21-W3 in the D21 
(first row) and the D2c (second row) difference images in the eight NIRCam 
bands as labeled. The images are 2\farcs4 on a side and oriented north-up, east-left. The red arrows mark the locations of transients
D21-W1, D21-W2, and D21-W3 from left to right in each image. 
D21-W1 (visible in all
epochs) is fainter in D2c (13.3 days apart) than in D21 (83.4 days apart), 
suggesting that it faded rapidly after reaching its maximum in Ep2. D21-W2
(not visible in Ep1) is almost invisible in Dc2, suggesting that it decayed 
more slowly. D21-W3 (not visible in Ep1) was similar to D21-W2 (and is only
$0\farcs28$ away) but fainter.
(bottom three rows, left to right) SEDs in each observable epoch, 
light curves in the SW bands, and light curves in the LW bands. D21-W1 was 
visible in all epochs, and the photometry was done on the original images.
The other two transients were not seen in Ep1, and their photometry was done on the 
difference images with respect to Ep1, i.e., the D21, Dc1 and D31 images. 
}
    \label{fig:W123}
\end{figure*}

\begin{figure*}[t]
  \epsscale{1.0}
  \plotone{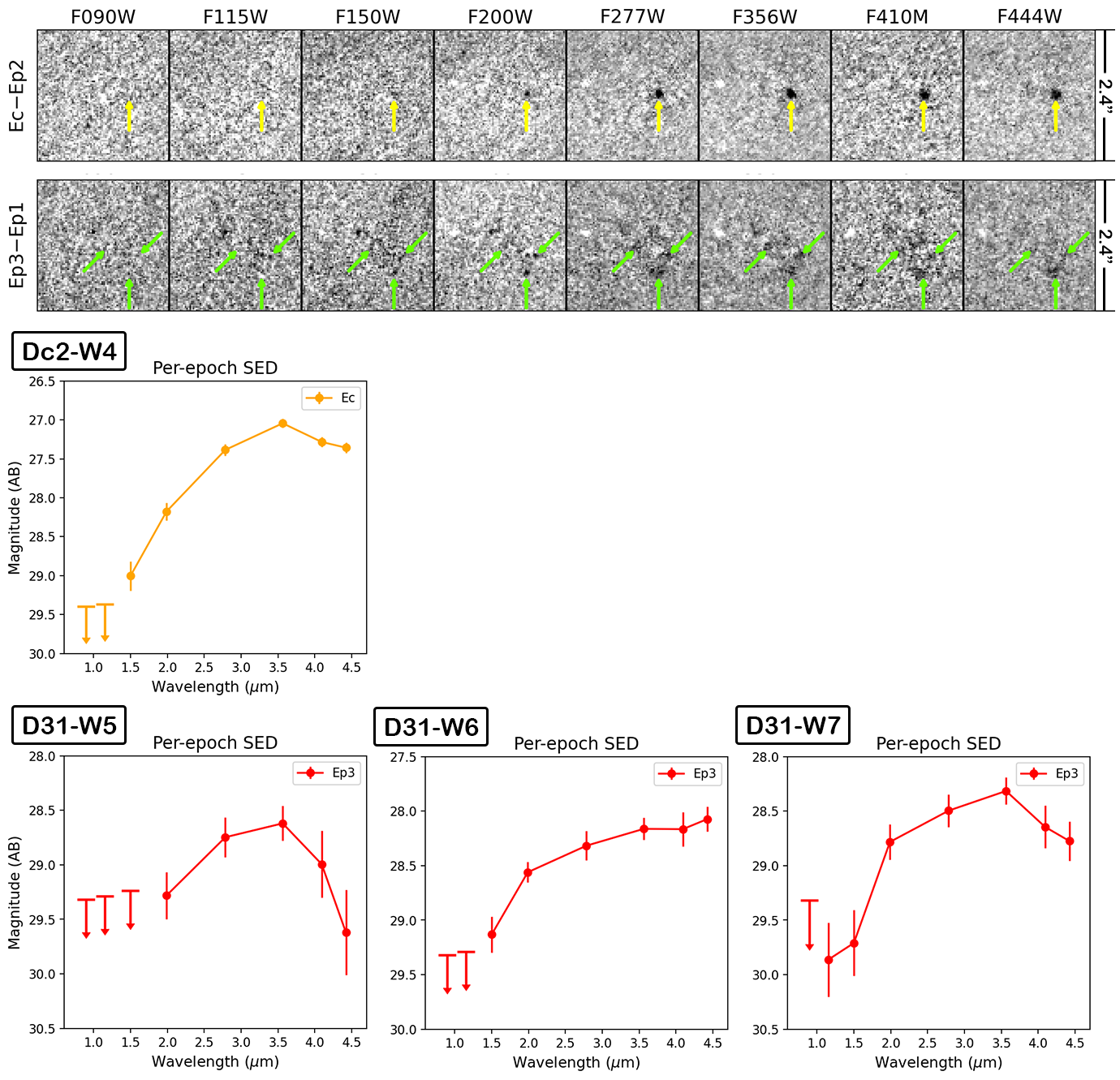}
    \caption{(top two rows) Transients Dc2-W4 in the Dc2 difference images
(first row) and D31-W5, D31-W6, and D31-W7 in the D31 difference images (second
row) in eight NIRCam bands. The images are 2\farcs4 on a side and oriented north-up, east-left. Dc2-W4 (position marked by the yellow arrow in each 
image) only appeared in Ec and was not visible in any other epoch, including Ep2 
only 13.3 days prior. D31-W5, D31-W6, and D31-W7 (positions marked by the 
green arrows from left to right in each image) were visible only in Ep3, and their detections
were significant only in the LW bands. 
(bottom two rows) SEDs of transients Dc2-W4 and D31-W5, D31-W6, and D31-W7.
}
    \label{fig:W4W567}
\end{figure*}

   Figure \ref{fig:WR} shows the positions of the seven transients discovered
in this region. The first three letters in their IDs indicate the difference
images on which they were first detected in our process; for example, ``Dc2''
stands for the difference image constructed by subtracting the Ep2 image from 
the Ec image, and so on. The letter ``W'' indicates that these are in the
Warhol region. The photometric results are listed in Table \ref{tbl:WR_phot}. 
Figure \ref{fig:W123} shows three of them (\texttt{D21-W1}, \texttt{D21-W2} and
\texttt{D21-W3}) that were seen in multiple epochs, while Figure 
\ref{fig:W4W567}
shows the other four (\texttt{Dc2-W4}, \texttt{D31-W5}, \texttt{D31-W6}, 
\texttt{D31-W7}) that were visible in only a single epoch.

   $\bullet$ \texttt{D21-W1}\,\,\, This transient was visible in Ep1, reached 
its maximum brightness in Ep2, became fainter in Ec, and further declined in 
brightness in Ep3 but remained visible. It faded more rapidly in the red 
bands than in the blue ones. At its peak ($m_{277}=27.10$~mag), it was the 
brightest among all transients in this region.  As it was visible in all 
epochs, its photometry was done in each epoch individually. 

   $\bullet$ \texttt{D21-W2}\,\,\, This transient was invisible in Ep1, 
appeared in Ep2, and slowly varied in the following two epochs. Interestingly, 
its behaviors in the SW and the LW bands  differed: while it decayed with time in
the LW bands, it became much brighter in the SW bands in Ep3, especially in the two
bluest bands. The photometry was done on the difference images between these 
epochs and Ep1 (i.e., the D21, Dc1 and D31 images), as this offers a more
reliable determination of the background. 

   $\bullet$ \texttt{D21-W3}\,\,\, This transient was only $0\farcs28$ away 
from \texttt{D21-W2} and was also invisible in Ep1. It appeared in Ep2 in F150W 
and redder bands. It was much weaker in Ec and barely (if at all) visible in 
Ep3. The photometry was done on the difference images between all other epochs 
and Ep1. The decline in brightness from Ep2 to Ec is very obvious in the blue 
bands. The F444W photometry weakly suggests that it might have slightly 
brightened from Ep2 to Ec, but this is inconclusive because of the large 
uncertainties. The extracted signals in Ep3 all have $\rm{S/N}< 2$, which we consider 
as non-detections. 

   $\bullet$ \texttt{Dc2-W4}\,\,\,  This event appeared as a sudden brightening 
in Ec, particularly in the LW bands. As mentioned earlier, Ec was not used to 
initiate the transient search; this event was found on the difference images
involving Ec when inspecting other transients in the Warhol region. While there 
seems to be a ``source'' in other epochs at this location, there is no
detectable signal in the difference images between Ep1, Ep2, and Ep3. This 
means that the event happened only in Ec, and it left no trace in any other 
epochs, including Ep2, which was only 13.3 days prior. The photometry was done 
on the difference images between Ec and Ep1 (i.e., the Dc1 images). 

   $\bullet$ \texttt{D31-W5}\,\,\, This transient was seen only in Ep3, as it 
was only visible in the difference images involving Ep3. It was very close to 
\texttt{D21-W3} but was a different transient. It was invisible in the three
bluest bands. The photometry was done on the difference images between Ep3 and 
Ep1 (the D31 images).

   $\bullet$ \texttt{D31-W6} and \texttt{D31-W7} \,\,\, These two transients
were also seen only in Ep3. Like \texttt{D31-W5}, the photometry was done on 
the difference images between Ep3 and Ep1 (the D31 images).

\begin{figure*}[t]
  \epsscale{1.2}
  \plotone{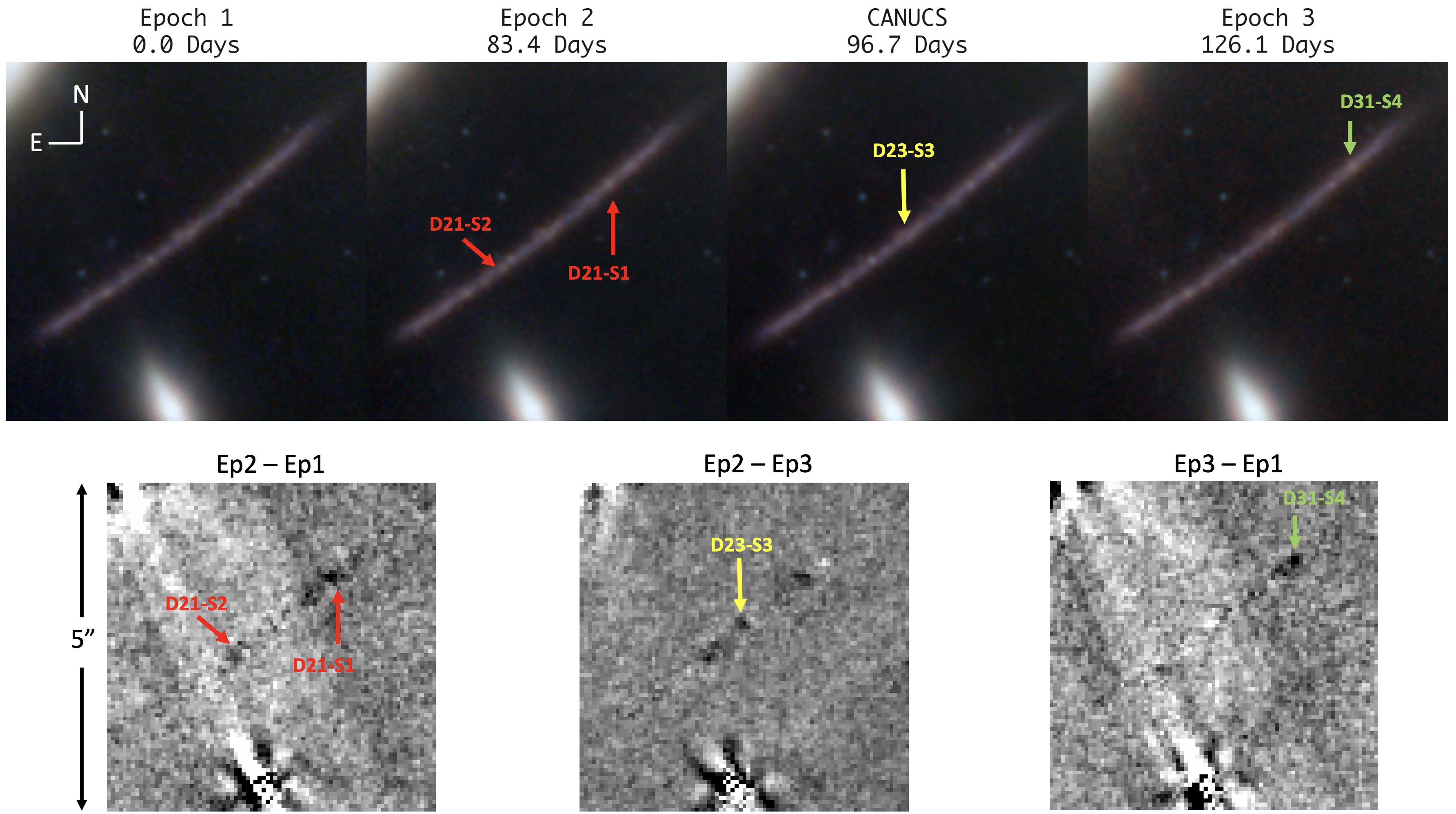}
    \caption{Locations of the transients found in the Spock region. The upper
panel shows the color images  of this region in the four epochs, while the 
lower panels show the inverted F356W difference images of the same region.  
 All images are 5\arcsec\ on a side and are oriented north-up 
and east-left.}
    \label{fig:SR}
\end{figure*}

\begin{table*}[hbt!]
\centering
\caption{Catalog of transients in the Spock arc region}
\label{tbl:SR_phot} 
\resizebox{0.99\textwidth}{!}{
\begin{tabular}{ccccccccccccc}\hline
 & R.A. & Decl. & Epoch & F090W & F115W & F150W & F200W & F277W & F356W & F410M & F444W & $\mu$ \\ \hline
 {\bf D21-S1} & 64.03847 & $-$24.06984 & Ep2 & $>$29.13 & $>$29.17 & $>$29.14 & $>$29.36 & 28.55\textpm0.11 & 28.56\textpm0.09 & 28.74\textpm0.17 & 29.00\textpm0.14 & 612.6\\ 
 & & & Ec & $>$29.17 & $>$29.21 & $>$29.24 & $>$29.44 & 29.17\textpm0.16 & 29.31\textpm0.17 & 29.06\textpm0.25 & 29.83\textpm 0.27 & \\ \hline
 
 {\bf D21-S2} & 64.03889 & $-$24.07017 & Ep2 & $>$29.13 & $>$29.17 & 29.65\textpm0.27 & 29.13\textpm0.17 & 28.33\textpm0.10 & 28.61\textpm0.12 & 29.12\textpm0.34 & 29.33\textpm0.27 & 87.7 \\ \hline
 
 {\bf D23-S3} & 64.03874 & $-$24.07004 & & & & & & & & & & 327.6 \\ \hline
 
 {\bf D31-S4} & 64.03836 & $-$24.06978 & Ep3 & $>$29.11 & 29.95\textpm0.29 & 29.86\textpm0.33 & 29.06\textpm0.17 & 28.55\textpm0.13 & 28.44\textpm0.09 & 28.68\textpm0.20 & 28.49\textpm0.11 & 139.0 \\ \hline
 
\end{tabular}
}
\raggedright
\tablecomments {As for Table \ref{tbl:WR_phot} but for the transients in the
Spock arc region. The lensing magnification factors $\mu$ are from
\citet[][]{Bergamini2023}.
}
\end{table*}

\subsubsection{Transients in the Spock region}

\begin{figure*}[t]
  \epsscale{1.1}
  \plotone{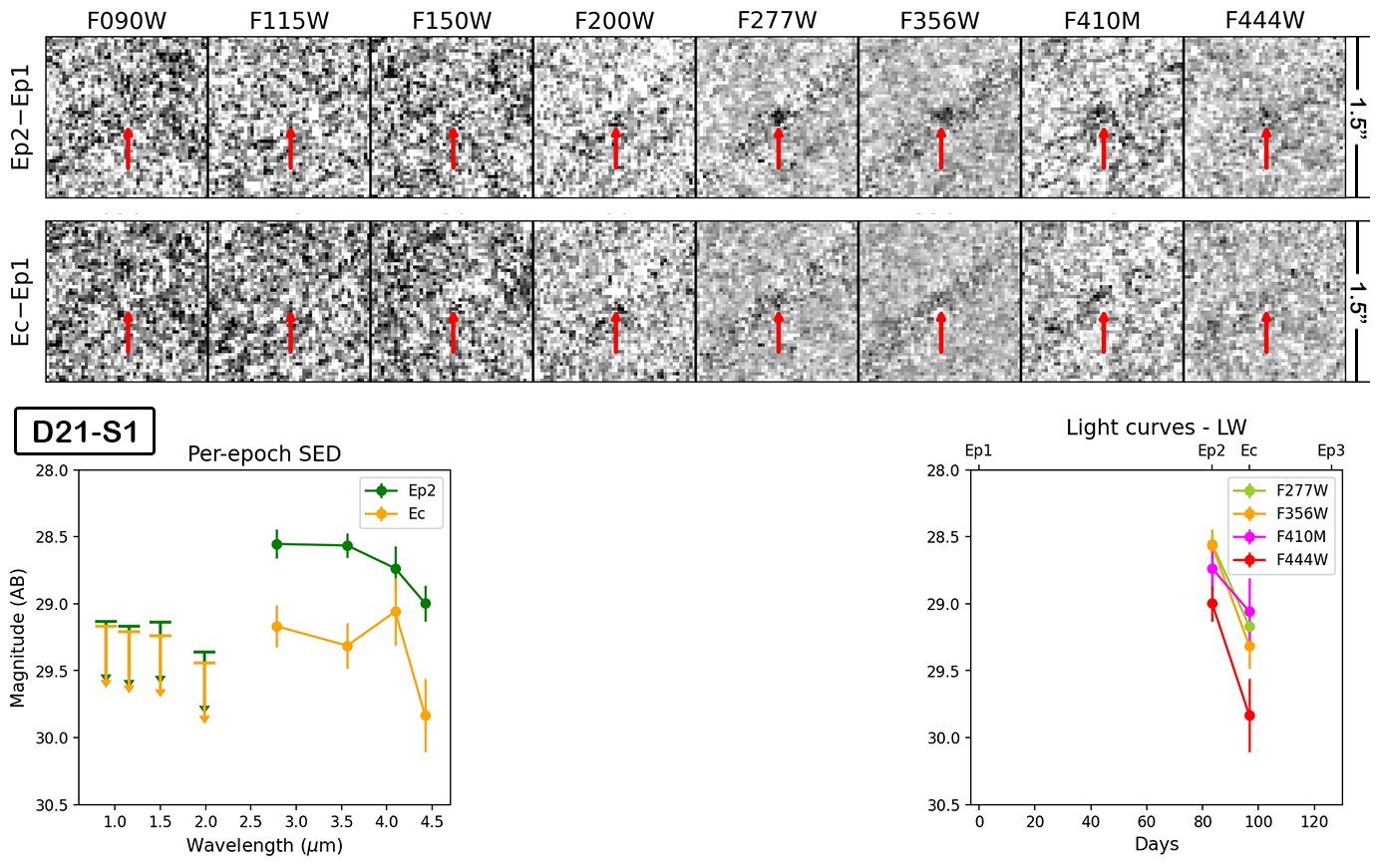}
    \caption{(top two rows) Transient D21-S1 in the D21 (first row) and Dc1 
(second row) difference images. The size of the images is labeled. The 
transient location is marked by the red arrow in each image. It is most 
prominent in the D21 images and becomes weaker in Dc1 images (but still 
visible), indicating that it reached maximum in Ep2 and then rapidly 
declined (disappearing from Ep3 entirely). It is almost invisible in the 
difference images in the SW bands.
(bottom row) Photometric information
of this transient, similar to those presented in Figure \ref{fig:W123} but
without the light curves in SW as it was not seen in these blue bands. The photometry in Ep2 and
Ec is based on the D21 and Dc1 difference images, i.e., it assumes the object was invisible in Ep1. 
}
    \label{fig:S1}
\end{figure*}

   Figure \ref{fig:SR} shows the positions of the four transients in this 
region. Due to the high brightness of this arc, these transients can only be 
revealed in the difference images, and none of them is clearly seen in the 
original images. Their IDs follow the convention of 
\S3.2.1 with ``S''  to indicate that these 
transients are in the Spock region. Figures~\ref{fig:S1} to~\ref{fig:S4} show
the details of these transients. Their photometry (except for 
\texttt{D23-S3}, see below) is presented in Table \ref{tbl:SR_phot}.

  $\bullet$ \texttt{D21-S1}\,\,\, This transient is best detected in the 
difference images between Ep2 and Ep1 (the D21 images) but is seen only in the 
LW bands. It became significantly weaker in the difference images between Ec 
and Ep1 (the Dc1 images) and almost completely disappeared from between 
Ep3 and Ep1 (the D31 images). All this indicates that it reached the maximum in 
Ep2 and then faded. Assuming that it was invisible in Ep1, we obtained its 
magnitudes in Ep2 and Ec by photometry on the difference images between Ep2 and 
Ep1 (D21) and those between Ec and Ep1 (Dc1).

  $\bullet$ \texttt{D21-S2}\,\,\, This transient was detected in the difference 
images involving Ep2 but not otherwise. Therefore, it is reasonable to assume 
that this event was caught in Ep2 only. The photometry was done on the 
difference images between Ep2 and Ep3 (i.e., the D23 images) because this combination offers 
a cleaner background than others (e.g., the D21 images).

\begin{figure*}[t]
  \epsscale{1.1}
  \plotone{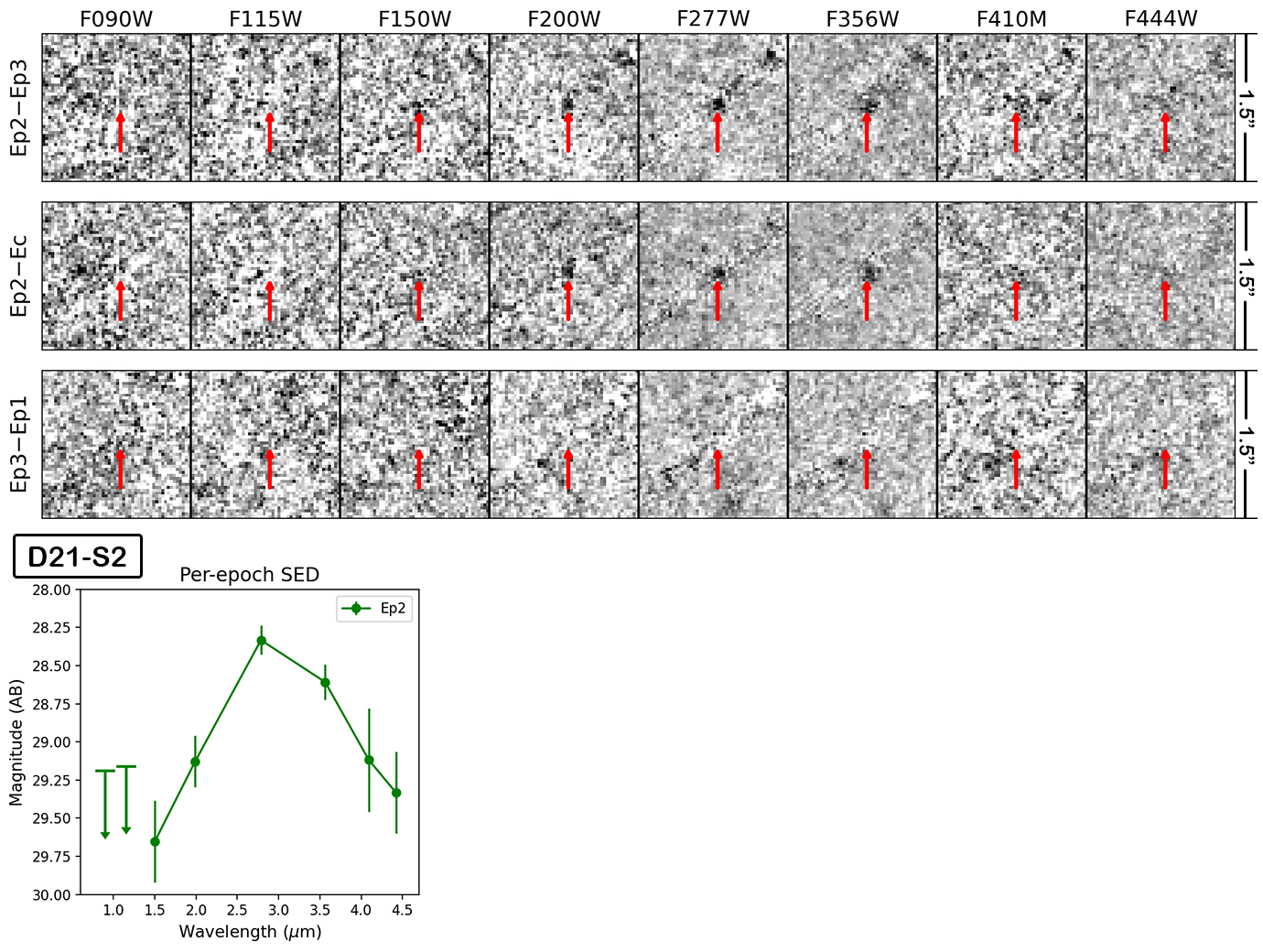}
    \caption{(top three rows) Transient D21-S2 in the D23 (first row), D2c
(second row), and D31 (third row) difference images (negative). The size of the images is
labeled.  The location of the transient is marked by the red arrow in each 
image. It is visible in the difference images involving Ep2 but not otherwise, 
indicating that the event was caught only in Ep2.
(bottom row) SED of the transient in Ep2 as measured from the D23 images.
}
    \label{fig:S2}
\end{figure*}

  $\bullet$ \texttt{D23-S3}\,\,\, This transient was best detected in the 
difference images between Ep2 and Ep3 (D23). It appears in the D23 images in 
F150W through F410M, is barely visible in F444W, and is invisible in F115W and 
F090W. This transient presents a complicated case that is difficult to 
understand. First of all, it seems to be an elongated system in the D23 F356W 
image. In the D23 F200W and F150W images, which have better resolution, this 
elongated structure is resolved into two components. However, it does not 
maintain such a two-component structure (or the elongated morphology) 
consistently in all bands: one of the components (the southern one)
is missing from the D23 F277W and F410M images.  Second, in the 
difference images between Ep2 and Ep1 (D21), only the southern component
appears, and it appears only in F200W and F150W\null. Taking the above at the face
value, one would infer the following picture: (1) D21-S3 was made of two
components; (2) the northern one maintained its brightness from Ep1 to Ep2 and
then decayed (not visible in the D21 images but showing up in the D23 images);
(3) the southern component brightened in Ep2 but only in F150W and F200W (in the
D21 images visible only in F150W and F200W) and then decayed; (4) however,
this southern component maintained its brightness from Ep1 through Ep3 in
F277W, F356W, and F410M\null. The last point is inconsistent with the observation 
that the southern component seems to be present in the D23 F356W image. It is
possible to attribute this inconsistency to the weakness of the signals.
We attempted PSF fitting on the two components in the D23 images, but the
fitting failed in all bands. We have to give up photometry on this transient.

\begin{figure*}[t]
  \epsscale{1.1}
  \plotone{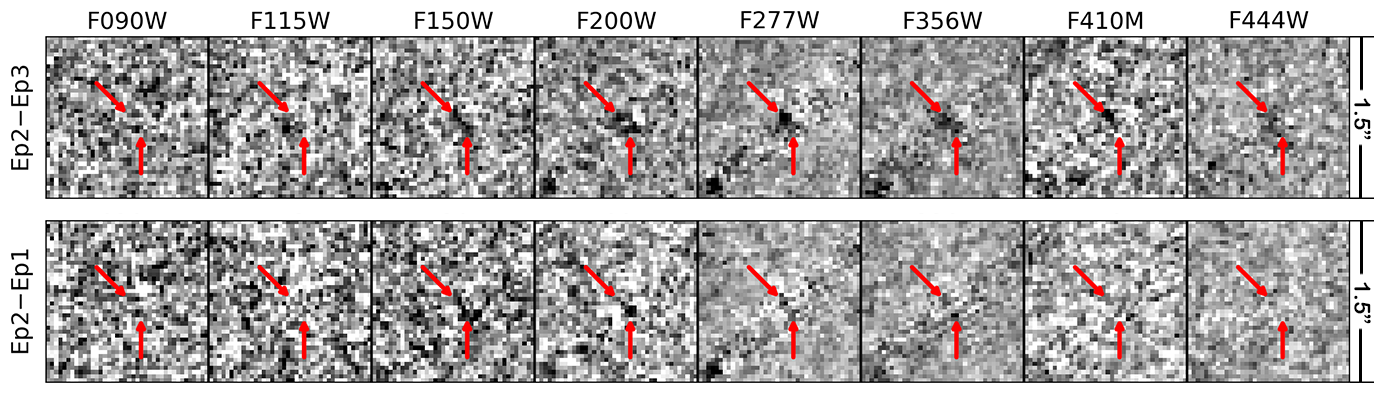}
    \caption{Transient D23-S3 in the D23 (first row) and D21 difference images.
The images have the same size as labeled. The transient location is marked by 
the red arrow in each image. This transient, however, is the most
complicated of all, and the difference images of different epoch pairs
reveal some inconsistencies. See text for details.
}
    \label{fig:S3}
\end{figure*}

  $\bullet$ \texttt{D31-S4}\,\,\,  This transient was only detected in the 
difference images involving Ep3, implying that it appeared in Ep3. It is only 0\farcs39 away from \texttt{D21-S1}, which already decayed and was
invisible in Ep3. The photometry was done on the difference images between Ep3 
and Ep1 (D31).

\begin{figure*}[t]
  \epsscale{1.1}
  \plotone{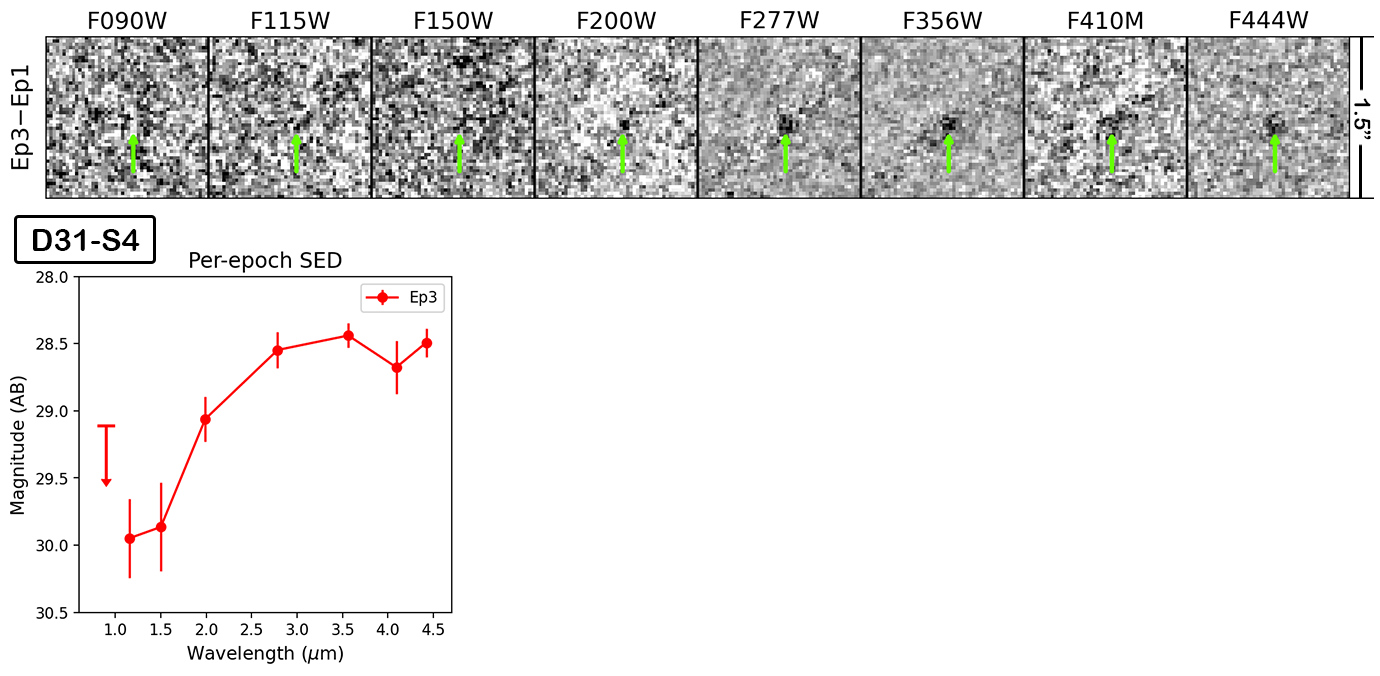}
    \caption{(top) Transient D31-S4 in the D31 (negative) difference images. The size of
the images is labeled.  The transient location is marked by the green arrow in
each image. It was only seen in the difference images involving Ep3, indicating 
that it appeared only in that epoch. (bottom) SED of this transient in Ep3 as measured on 
the D31 difference images.
}
    \label{fig:S4}
\end{figure*}

\begin{table*}[hbt!]
\centering
\caption{Photometry for the transient ``Mothra'' in an arc at $z=2.091$.}
\label{tbl:YAA_phot}
\resizebox{0.95\textwidth}{!}{
\begin{tabular}{cccccccccccc}\hline
  R.A. & Decl. & Epoch & F090W & F115W & F150W & F200W & F277W & F356W & F410M & F444W & $\mu$ \\ \hline
 64.03676 & $-$24.06625 & Ep1 & 28.19\textpm0.19 & 28.12\textpm0.17 & 28.11\textpm0.24 & 28.07\textpm0.25 & 27.96\textpm0.24 & 27.96\textpm0.25 & 27.89\textpm0.23 & 27.88\textpm0.22 & 32.5 \\ 
  & & Ep2 & 28.19\textpm0.19 & 28.12\textpm0.17 & 28.11\textpm0.24 & 27.92\textpm0.23 & 27.77\textpm0.21 & 27.46\textpm0.16 & 27.47\textpm0.17 & 27.49\textpm0.16 & \\ 
  & & Ec & 28.19\textpm0.19 & 28.12\textpm0.17 & 27.92\textpm0.21 & 27.66\textpm0.18 & 27.46\textpm0.16 & 27.37\textpm0.15 & 27.29\textpm0.14 & 27.27\textpm0.13 & \\
  & & Ep3 & 28.19\textpm0.19 & 28.12\textpm0.17 & 27.88\textpm0.20 & 27.76\textpm0.20 & 27.40\textpm0.15 & 27.36\textpm0.15 & 27.24\textpm0.14 & 27.24\textpm0.13 & \\ \hline
\end{tabular}
}
\raggedright
\tablecomments{As for Table \ref{tbl:WR_phot} but for the transient in the
``Mothra'' arc region. 
The lensing magnification factor $\mu$ is from \citet[][]{Bergamini2023}.
}
\end{table*}

  Detecting four transients in this arc in the four
epochs is consistent with expectations for this particular arc. 
\cite{Diego2023a} found that microlensing should produce 
between one and five transients per pointing in the Spock arc when reaching
$\sim$29~mag.

\subsubsection{A transient in yet another arc}

\begin{figure*}[t]
  \epsscale{1.0}
  \plotone{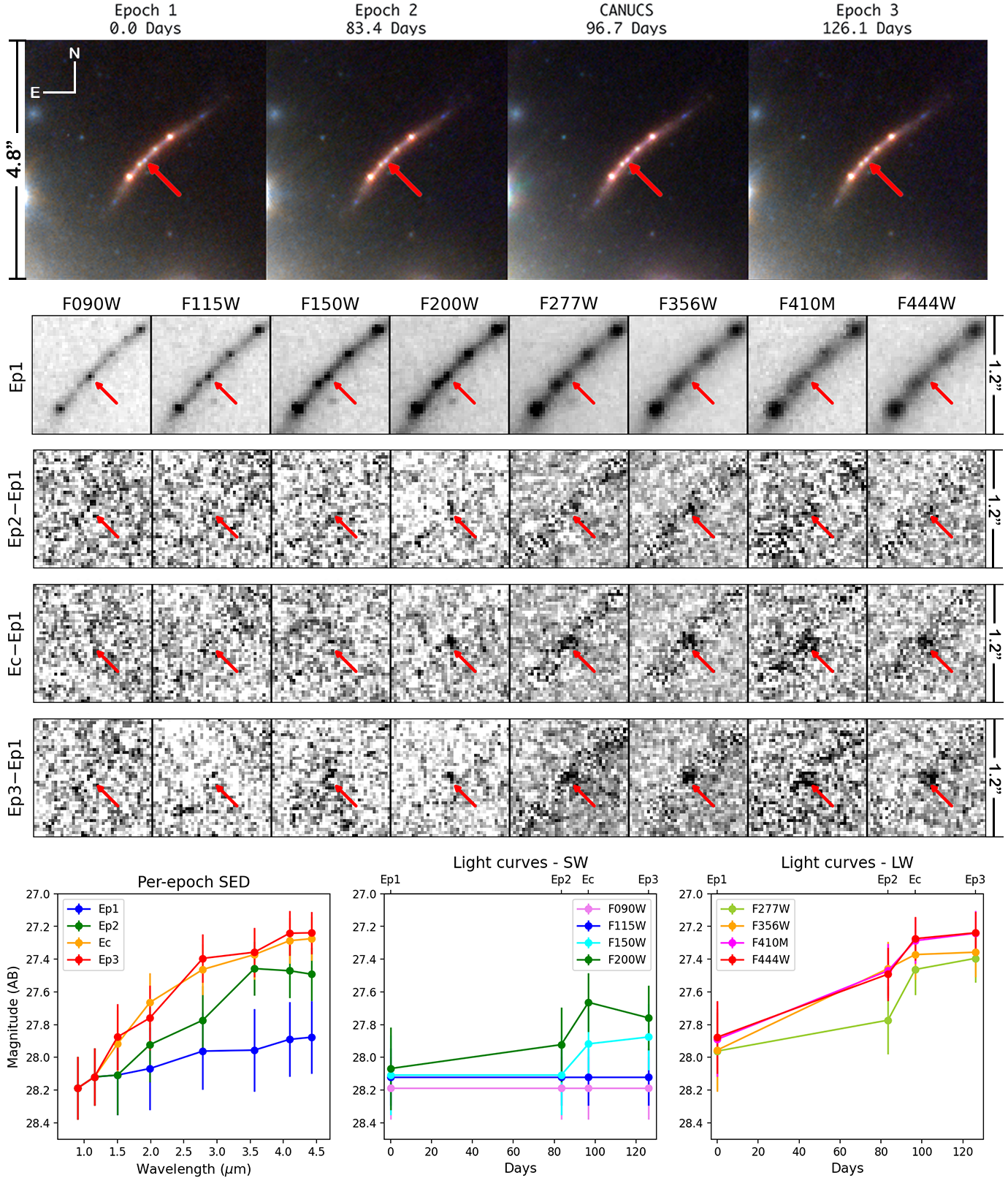}
    \caption{``Mothra'' transient identified in an unnamed arc. The top row 
shows the color images (4\farcs8 on a side) of this region in four epochs. The 
red arrow in each image points to the knot that gave rise to the event. The
second row zooms in to the arc and shows the Ep1 images (1\farcs2 on a side) 
in the eight bands. The next three rows show the difference images between 
other epochs and Ep1, all displayed in negative. The last row shows the 
photometric information. See \S3.2.3 and the Appendix for details of the 
photometry.
}
    \label{fig:yaa}
\end{figure*}

   There is one transient identified on an arc at $z=2.091$ 
\citep[][]{Bergamini2021} where no previous transient has been 
reported. Dubbed ``Mothra,'' this transient is discussed in detail by 
\citet[][]{Diego2023b}.
Figure~\ref{fig:yaa} shows the details of this transient. The knot in which it 
is located is the faintest among five knots on the arc, but the knot  was 
visible in all four epochs.  This transient is best explained by the intrinsic 
variability of a red supergiant star (in a binary system with a blue 
supergiant) that is being magnified by a dark milli-lens \citep[][]{Diego2023b}.

   The transient behavior is best seen in the difference images between Ep3
and Ep1 (D31) and as well as in those between Ec and Ep1 (Dc1), where it shows 
up as a strong, red source  with decreasing amplitude towards the blue end. It 
is even visible in the difference images between Ec and Ep2 (Dc2; 13.3 days 
apart), albeit being much weaker. It is almost invisible in the difference 
images between Ep3 and Ec (D3c; 29.4 days apart). In other words, this knot 
was slowly increasing in brightness and reached maximum at around Ec (96.7 
days between Ep1 and Ec), and it stayed more or less at its maximum through 
Ep3 (29.4 days between Ec and Ep3).

   Because this transient was caused by the variability of a source that
was visible in all epochs, ideally its photometry should be done on the images
taken in each epoch. As it is almost blended with a brighter knot 
0\farcs12 to the southeast, one should do PSF fitting on both simultaneously. 
However, we were only
able to obtain reasonable PSF fitting results in Ep1. (The Appendix gives 
details.) The procedure failed in other epochs, mostly because the light of 
brightened transient blended with the nearby knot more severely. Therefore, we 
performed PSF fitting on the difference images between Ep1 and other epochs 
(Figure~\ref{fig:yaa}) and then added the excess fluxes extracted in this 
way to the Ep1 SED to obtain the SEDs in other epochs. The results are 
presented in Table~\ref{tbl:YAA_phot} and are also shown in the bottom row of 
Figure~\ref{fig:yaa}.

\begin{table*}[hbt!]
\centering
\caption{Photometry for the two SNe}
\label{tbl:SNe_phot} 
\resizebox{0.95\textwidth}{!}{
\begin{tabular}{cccccccccccccc}\hline
 & R.A. & Decl. & Epoch & F090W & F115W & F150W & F200W & F277W & F356W & F410M & F444W & $z$ & $\mu$ \\ \hline
 {\bf SN01} & 64.02954 & $-$24.09022 & Ep1 & 26.84\textpm0.05 & 26.85\textpm0.04 & 27.25\textpm0.03 & 27.73\textpm0.04 & 28.29\textpm0.07 & $>$29.82* & $>$29.24* & $>$29.49* & 2.205 & 2.9 \\ 
 & & & Ep2 & 30.70\textpm0.41 & 29.14\textpm0.10 & 27.68\textpm0.04 & 27.16\textpm0.02 & 27.36\textpm0.04 & 27.47\textpm0.04 & 27.55\textpm0.05 & 27.64\textpm0.06 & & \\
 & & & Ec & $>$28.73* & 29.76\textpm0.23 & 27.98\textpm0.04 & 27.30\textpm0.03 & 27.56\textpm0.04 & 27.55\textpm0.04 & 27.55\textpm0.06 & 27.72\textpm0.07 & & \\ 
 & & & Ep3 & $>$28.87* & 29.80\textpm0.24 & 28.17\textpm0.05 & 27.42\textpm0.03 & 27.63\textpm0.04 & 27.74\textpm0.05 & 27.78\textpm0.07 & 27.96\textpm0.08 & & \\ \hline
 
 {\bf SN02} & 64.04421 & $-$24.07449 & Ep2 & 28.77\textpm0.11 & 27.85\textpm0.06 & 27.44\textpm0.05 & 27.31\textpm0.04 & 27.35\textpm0.06 & 27.60\textpm0.08 & 27.52\textpm0.09 & 27.68\textpm0.08 & 0.793 & 1.9 \\
 & & & Ec & 29.20\textpm0.16 & 28.09\textpm0.06 & 27.62\textpm0.04 & 27.39\textpm0.04 & 27.50\textpm0.07 & 27.71\textpm0.07 & 27.85\textpm0.11 & 28.07\textpm0.11 & & \\ 
 & & & Ep3 & 29.71\textpm0.31 & 28.04\textpm0.09 & 27.76\textpm0.06 & 27.51\textpm0.06 & 27.51\textpm0.08 & 27.54\textpm0.08 & 27.52\textpm0.08 & 27.88\textpm0.10 & & \\ \hline
 \end{tabular}
}
\raggedright
\tablecomments {As for Table \ref{tbl:WR_phot} but for the two SNe in less
magnified regions. The lensing magnification factors $\mu$ are from
\citet[][]{Bergamini2023}.
}
\end{table*}

\subsubsection{Two likely supernovae}

   There were two transients associated with galaxies that are only moderately
magnified. Both transients were detected in multiple epochs, and neither was
seen in the HFF data. Based on their light curves, we believe that they are 
SNe. Their physical interpretations will be detailed in a forthcoming paper 
(Wang et al., in prep.). 
Their photometry is presented in Table~\ref{tbl:SNe_phot}.

\begin{figure*}[t]
  \epsscale{1.2}
  \plotone{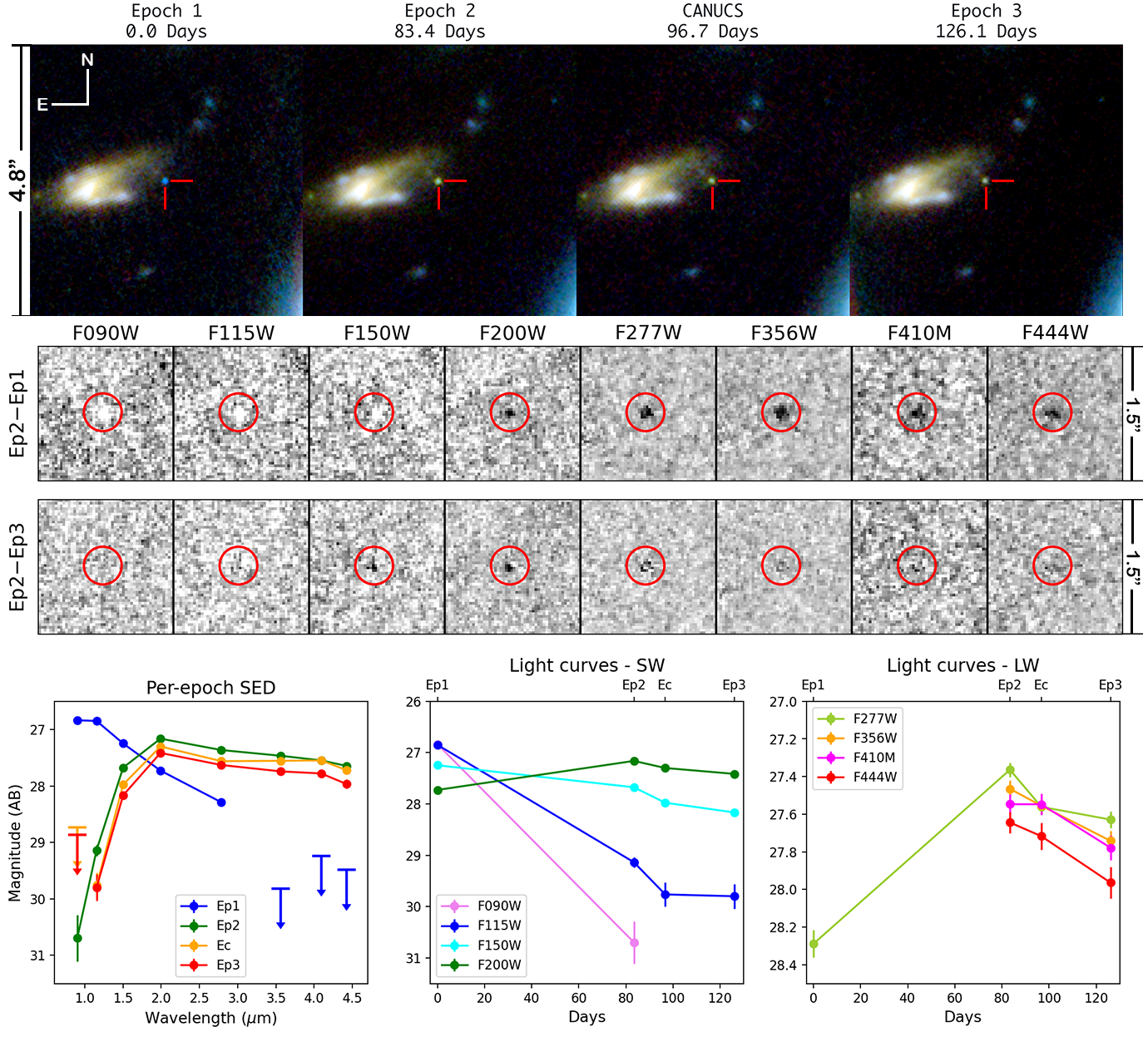}
    \caption{(top row) Color images of SN01 in four epochs. Scale and
orientation are labeled. The position of the SN is indicated by the red arrow 
in each image. The nearby irregular galaxy, which is at $z=2.205$, is 
presumably the host. 
(middle two rows) D21 and D23 (negative) difference images. The 
location of the source is indicated by the red circle (0\farcs5 radius) in 
each image. The change of color from blue to green/yellow is obvious in the 
D21 images, where its location shows negative signals in the three bluest 
bands and positive signals in the rest. It largely maintained the same color 
from Ep2 through Ep3 (although becoming dimmer) as seen in the D23 images. 
(bottom row) SED evolution over four epochs (left panel) and the light curves 
in the SW bands (middle) and the LW bands (right). In Ep1, this transient was 
invisible in the three reddest bands, and the downward arrows in its Ep1 SED
indicate the 5$\sigma$ upper limits (calculated on the RMS maps within a 
0\farcs165-radius circular aperture to best match the size within 
which the PSF fitting was done; see the Appendix). In F200W and the redder 
bands, the transient reached the maximum in Ep2 and then gradually faded.
}
    \label{fig:sn01}
\end{figure*}

   $\bullet$ \texttt{SN01}\,\,\, We initially reported this event 
\citep{Yan2023TNSa} based on the data from Ep1 and Ep2. Figure \ref{fig:sn01} 
shows the color images of the transient and its vicinity in the four epochs. 
The transient appeared as a blue source in Ep1 and then became very red in the 
subsequent epochs. The difference images show that its F200W and redder light 
reached  maximum in Ep2. The source is very close to an irregular galaxy, 
which presumably is the host. The CANUCS NIRISS slitless spectroscopy shows 
that this galaxy is at $z=2.205$ (C.\ Willot, private communication).
Because the transient was visible in all epochs, its 
photometry should be done on the images in individual epochs. To minimize the 
impact of the contamination from the host galaxy, the photometry was done by 
PSF fitting. The results are  shown in Figure~\ref{fig:sn01}.

   $\bullet$ \texttt{SN02}\,\,\, This transient was found within a spiral 
galaxy identified at $z=0.7093$ (redshift based on \citealt{Caminha2017}). 
The transient was invisible in Ep1 
and appeared in Ep2. From the difference images, this transient was brightest 
in most bands in Ep2 and then slightly decayed in Ec and Ep3. The photometry 
in Ep2, Ec, and Ep3 was done on the difference images between Ep2 and Ep1 
(D21), Ec and Ep1 (Dc1), and Ep3 and Ep1 (D31). In these difference images, 
the transient's neighbourhood is affected by the strong residuals from 
imperfect subtraction of the host-galaxy bulge. PSF-fitting photometry reduced 
but did not eliminate the contamination.

\begin{figure*}[t]
  \epsscale{1.2}
  \plotone{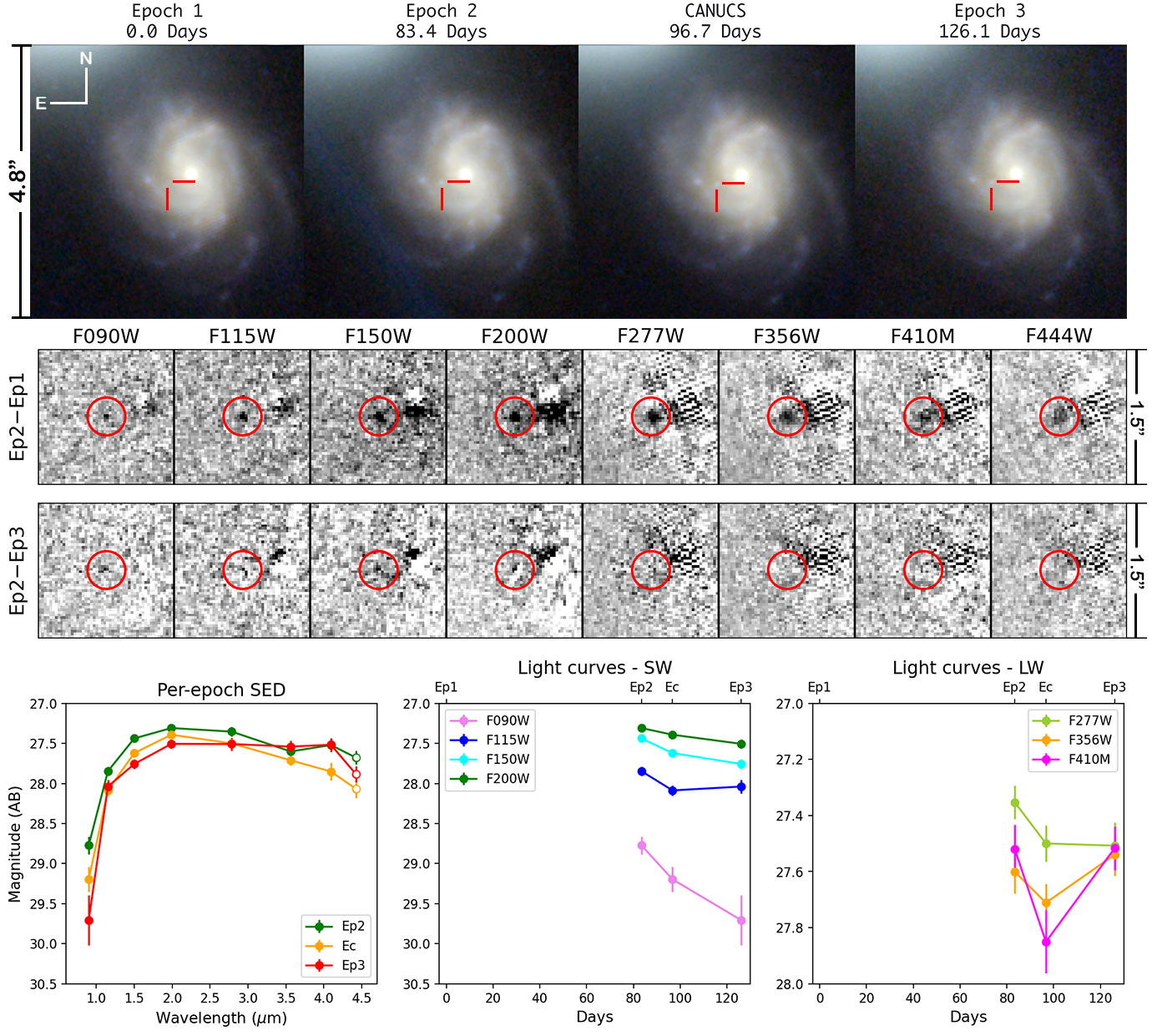}
    \caption{(top row) Color images of SN02 in four epochs. Scale and
orientation are labeled.  The two red bars in each of the color images 
indicate SN02's location. It was not visible in Ep1. The host redshift is 
$z=0.7093$.
(middle two rows) D21 and D23 (negative) difference images. The 
location of the source is indicated by the red circle (0\farcs5 radius) in 
each image. From the difference images involving Ep2, it seems that this 
transient reached the maximum in Ep2 and then faded.
(bottom row)
SED evolution over four epochs (left panel) and the light curves in the SW
bands (middle) and the LW bands (right).
}
    \label{fig:sn02}
\end{figure*}

\section{\label{discussion} Discussion}

    The NIRCam data used here were produced by only half of the NIRCam field of
view (module~B). Because of the different field orientations at different
times, the area overlapped in all four epochs amounts to only 3.98~arcmin$^2$.
Those data resulted in 14 robust transients, the largest number of transients 
ever found within such a small area. There are two reasons for this high 
transient production rate. First, the high sensitivity of NIRCam allows
searching for transients to an unprecedented depth. The vast majority of our 
transients were fainter than 27.0~mag even at their peaks and were fainter 
than 28.0~mag most of the time in most bands. The PEARLS observations had 
exposure times of $\sim$0.8--1 hours per band per epoch, and the data reach 
5$\sigma$ limits of $\sim$28.5--30.0~mag (within a 0\farcs2-radius aperture). 
This is deep enough to validate the transients in multiple bands. Second, 
M0416 includes two regions extremely magnified by the cluster, the Warhol and 
the Spock arcs, which are known to have produced a number of caustic 
transients in the previous studies with HST\null. Both arcs are at relatively 
low redshifts ($z\approx 1$), which facilitates the detection of luminous 
stars in them. In our data spanning 126 days, these two regions contributed 
seven and four transients, respectively, making them the most productive 
transient ``factories'' known. In addition, our search found a transient in 
another arc (the ``Mothra'' arc) where no transient had been seen previously. 
As mentioned above (and to be discussed in detail in  
forthcoming papers), these transients are most likely stars in the lensed arcs 
that were temporarily magnified by an extra factor by micro-lensing. 
Studying such transients remains the only direct way to study individual stars
at cosmological distances and therefore should be pursued by JWST in the 
coming years. These transients might occur very frequently. For example, 
\texttt{Dc2-W4} in the Warhol region appeared in Ec but was not seen in either 
Ep2 (13.6 days earlier) or Ep3 (29.4 days later). A monitoring cadence of 
$\sim$10 days ($\sim$5 days in the rest frame) would potentially reveal more 
fast transients like \texttt{Dc2-W4}.  This possibility is worth exploring 
with JWST in the near future.

    In addition to the 12 transients in the highly magnified arcs, we also 
discovered two transients in regions that are only moderately magnified
($\mu \sim 2$--3). They are most 
likely SNe, and with intrinsic, post-peak brightnesses of $\sim$28.5~mag,
they were bright enough to have been discovered even without the lensing 
magnification. Taken at face value, discovering two SNe in 3.98~arcmin$^2$ 
implies a SN surface density of $\sim$0.5~arcmin$^{-2}$ integrated to 
$z\approx 2.2$ when monitored over $\sim$126~days. This frequency is broadly 
consistent with expectations \citep[][]{WangFLARE2017,RV2019}. Both SNe were 
caught near their maxima (in the rest-frame visible range) by the Ep2 
observations, which were taken 83.4 days after Ep1. Owing to time dilation, 
neither transient changed significantly in brightness from Ep2 to Ec, which 
was 96.7 days after Ep1. This suggests that a time cadence of $\sim$90 days 
should be effective in discovering SN-like transients (integrated over all 
redshifts) and is likely to catch events near their peaks.

    Finally,  there could be a color bias in the transients reported here. The 
vast majority of them are very red with the only exception being SN01 in Ep1. 
Even that source transformed to a red object in Ep2. This may, at least in 
part, result from the initial selection being based on the F356W difference 
images. An initial selection based an SW band (e.g., F150W) is possible, 
although it would be more complex to validate because of the more numerous 
defects in the SW data. 

\section{\label{summary} Summary}

   M0416 has been observed by NIRCam for four epochs, making it the field most
intensely monitored by JWST in its Cycle~1. The eight-band data also provide 
the best near-IR SED sampling to date. This work has identified 14 transients 
in these four epochs, which spanned 126 days. Twelve transients occurred in 
three regions highly lensed by the cluster (seven, four, and one in the Warhol, 
Spock, and Mothra regions, respectively), while the other two happened in two 
background galaxies that are only moderately magnified (by $\sim$2--3$\times$). 
The eight-band photometry enables the construction of the transients' SEDs from 
0.9 to 4.4~$\mu$m. This is the first time that time-domain studies have such 
detailed information for interpretation. Further analysis of SEDs and light 
curves will be presented in forthcoming papers.

   This work demonstrates the power of JWST in the study of the transient IR
sky. It is now expected that JWST will be able to function for about twenty
years, enabling long-term monitoring programs addressing new science never 
before possible. A new era of IR time-domain science has begun.

The NIRCam data presented in this paper can be accessed via
\dataset[10.17909/wmmd-ev74]{http://dx.doi.org/10.17909/wmmd-ev74}
after the respective proprietary periods.

\begin{acknowledgments}
This work is dedicated to the memory of our dear colleague Mario Nonino, a
kind and gentle person and an example for many.
We thank the CANUCS team for generously providing early access to their
proprietary data of MACS0416. 
This project is based on observations made with the NASA/ESA/CSA James Webb 
Space Telescope and obtained from the Mikulski Archive for Space Telescopes,
which is a collaboration between the Space Telescope Science Institute 
(STScI/NASA), the Space Telescope European Coordinating Facility (ST-ECF/ESA), 
and the Canadian Astronomy Data Centre (CADC/NRC/CSA). We thank our Program 
Coordinator, Tony Roman, for his expert help scheduling this complex program.
This research made use of Photutils, an Astropy package for detection and 
photometry of astronomical sources.
HY and BS acknowledge the partial support from the University of Missouri 
Research Council Grant URC-23-029. 
JMD acknowledges the support of project PGC2018-101814-B-100 
(MCIU/AEI/MINECO/FEDER, UE) Ministerio de Ciencia, Investigaci\'on y 
Universidades. 
LW acknowledges support from NSF through grant \#1813825.
SHC, RAW, and RAJ acknowledge support from NASA JWST Interdisciplinary Scientist 
grants NAG5-12460, NNX14AN10G, and 80NSSC18K0200 from GSFC\null. 
ZM is supported in part by National Science Foundation grant \#1636621.
JFB was supported by NSF Grant No.\ PHY-2012955.
CNAW acknowledges funding from the JWST/NIRCam contract NASS-0215 to the
University of Arizona.
CC is supported by the National Natural Science Foundation of China, No.\ 
11803044, 12173045.
AZ acknowledges support by Grant No.\ 2020750 from the United States--Israel 
Binational Science Foundation (BSF),  Grant No.\ 2109066 from the United 
States National Science Foundation (NSF), and by the Ministry of Science 
\& Technology, Israel.
\end{acknowledgments}

\bibliographystyle{aasjournal.bst}

\appendix

   While we used isophotal aperture photometry by SExtractor to search for
transients, we used PSF fitting for the final photometry of the transients identified. 
This approach assumes that transients are all point-like even at 
the JWST resolution, which should be valid. The reason  we adopted the
more complicated PSF fitting (as opposed to aperture photometry) was because of background contamination. Our transients were all embedded in 
highly non-uniform background, and in a lot of cases that still left 
structures even in the difference images between epochs. In such situations, 
PSF fitting handles contamination better than any aperture photometry.
We did PSF fitting on the 30mas images (as opposed to the 60mas used in the transient identification). The process is outlined below.

    We first generated the PSFs using the simulation tool {\tt WebbPSF} 
(version 1.1.1) at the pixel scale of 30mas. For each band, PSFs were simulated
at 36 evenly distributed positions on each detector, and each simulated PSF was 
saved as an individual image. All simulated PSFs were 87$\times$87 pixels 
(2\farcs61$\times$2\farcs61) in size. Then, effective PSFs, referred to as
the WPSFs, were built using 
{\tt EPSFBuilder} in {\tt Photutils}. We also constructed empirical PSFs for comparison. 
The difficulty with those is that the M0416 field does not have many suitable stars. 
Nevertheless, we were able to find five isolated, unsaturated, and high 
${\rm S/N}$ stars for this purpose. A region of 87$\times$87 pixels centered
on each star was cut out from the image, and the five cutouts were sent to
{\tt EPSFBuilder} to build the empirical PSF,
referred to as the EPSFs. Both types of PSFs were used, and we found
only marginal differences. (See below.)

    We used the {\tt BasicPSFPhotometry} function in {\tt Photutils} 
\citep[][]{larry_bradley_2023} to perform PSF
fitting. This function allows simultaneous fitting to multiple, overlapping 
sources when necessary. The non-linear fitting routine {\tt LevMarLSQFitter} in
{\tt Astropy} \citep[][]{astropy2013, astropy2018, astropy2022}
was applied. The routine utilizes least-squares statistics to decide
on the best fit. In PSF fitting, the actual area used to fit the model is
usually much smaller than the full PSF size, as only the central region has
sufficient ${\rm S/N}$. Here we found that the optimal fitting area was 
11$\times$11 pixels, which is about 2.8$\times$ the full-width-at-half-maximum 
of the F356W~PSF. 

    We provided an initial guess of the source's centroid and flux
when running the fit. The former was estimated by visually locating the peak
pixel, while the latter was estimated by measuring the aperture flux within a
circle of 11 pixels in diameter centered at the initial guess of the location.
Tests showed that {\tt BasicPSFPhotometry} could converge to the same solution 
even when the initial guesses were widely different. On the other hand, the
routine requires an accurate background estimate because it fixes the background
to the input value. In most cases, we estimated the background by using the
{\tt MedianBackground} function in  {\tt Photutils} and adopting the 
3$\sigma$-clipped median in the image cutouts centered on the transient 
(i.e., the same image stamps as shown in the figures in the main text) with the
sources masked.

    Some special cases required tailored treatments of the fitting. For example, leaving both the source 
location and its flux as free parameters was not feasible for sources of
low ${\rm S/N}$. In these cases, we fixed the centroid and fit for the flux. For
several sources in highly magnified regions, small but significant positional 
offsets were found between some SW and LW bands. In such cases, we did not force
the fit to be centered at the same position; instead, we determined the 
centroid in different bands individually. As to the background estimate,
median statistics were not applicable for sources in extremely non-uniform
local backgrounds. For these cases, we estimated the local background value 
manually in an iterative manner: we subtracted different constants from the 
image at a step size of 0.005 MJy/sr, performed a PSF fit, and visually 
examined the residual image to determine the best local background value by
eye. Sources that required some of these special 
treatments were:
\begin{itemize}

    \item D21-W2\,\,\, We determined the source's centroids in the LW bands 
using the D21 images and in the SW bands using the D31 images. Their centroids
in each band were then fixed for all epochs. 

    \item D21-W3\,\,\, The source's centroids were determined from the D21
images and then fixed for Ec. 

    \item Dc2-W4\,\,\, This source's local background is non-uniform, and we 
visually examined and selected the best local background value in Ec. 

    \item D31-W5/W6/W7\,\,\, These sources' centroids in the SW bands were 
determined from the D31 F200W image. In the LW bands, they were determined from
the D31 F356W image. 

    \item D21-S2\,\,\, This source's centroids in the SW bands were determined 
from the F200W D21 image. In the LW bands, they were determined from the D21 
F277W image. 

    \item D31-S4\,\,\, This source's centroids in F115W and F150W were fixed to
its F200W centroid. 

    \item Mothra\,\,\, (1) In Ep1, this source's centroid in F444W was fixed 
to its F410M centroid. (2) The complexity of simultaneously fitting two 
overlapping sources on a thin arc made it hard to use any automatic approach
to estimate the local background. Therefore, we had to tune the background
estimate manually as mentioned above. Furthermore, we used the Ep1 difference 
images between adjacent bands (the bluer image was always PSF-matched to the 
redder image when constructing the difference) to judge whether the extracted 
flux in each band was reasonable under each step of background estimate. For 
example, if the F150W\,$-$\,F115W difference image showed a distinct source at the 
transient location, it meant that the extracted flux in F150W must be higher
than that in F115W\null. This added constraint, while tedious, allowed us to
tweak the background values to obtain the most reasonable flux measurements.

\end{itemize}

   In all cases, our PSF-fitting results have been properly normalized by
aperture correction. As mentioned above, we used five stars to construct the
EPSFs; these five stars were our basis for the aperture correction. We ran
PSF fitting on these five stars to obtain their fitted fluxes and also derived
their aperture fluxes within the same 11$\times$11 pixel areas. The averaged
ratio between the two in each band was the multiplicative aperture correction
factor, which we applied to the outputs from {\tt BasicPSFPhotometry}.

   In the end, there are only marginal differences between the 
results based on the WPSFs and those based on the EPSFs. As the EPSFs were
derived using only a small number of stars (total of five), we regard
them as being less secure. Therefore, we adopted the WSPF results for our final 
photometry reported in the tables.

\end{document}